\documentclass[11pt,superscriptaddress,notitlepage,longbibliography]{revtex4-1}
\usepackage{amsmath}
\usepackage{amssymb}
\usepackage{amsthm}
\usepackage{graphicx}
\usepackage{hyperref}
\usepackage{color}
\newcommand{\p}{\partial}
\begin{document}
\title{Modeling unsteady B\'enard-Marangoni instabilities in drying volatile
droplets on a heated substrate}
\author{A.A. Gavrilina}
\affiliation{National Research University Higher School of Economics, 101000 Moscow, Russia}
\affiliation{Landau Institute for Theoretical Physics, 142432 Chernogolovka, Russia}
\author{L.Yu. Barash}
\affiliation{Landau Institute for Theoretical Physics, 142432 Chernogolovka, Russia}
\begin{abstract}
We study unsteady internal flows in a sessile droplet of capillary size evaporating in 
constant contact line mode on a heated substrate. Three-dimensional simulations of internal 
flows in evaporating droplets of ethanol and silicone oil have been carried out. For 
describing the Marangoni flows we find it 
necessary to account for the diffusion of vapor in air,
the thermal conduction in all three phases and thermal radiation.
The equations have been solved numerically by finite element 
method using ANSYS Fluent. As a result of the simulations, the nonstationary behavior of 
B\'enard-Marangoni (BM) instabilities is obtained. At the first stage, a flower structure of 
BM cells near the triple line emerge. For smaller contact angles, the cells grow in size and 
occupy the central region of the droplet surface. 
Being closely connected with recent experimental and theoretical 
studies, the results obtained help to analyze and resolve the associated issues.
\end{abstract}
\maketitle

\section{Introduction}

The issues of droplet evaporation have attracted significant attention both from the 
theory point of view and in connection with the development of new applications, including 
pattern formation on surfaces, preparation of ultra-clean surfaces, deposition of 
microarrays of DNA and RNA molecules, protein crystallography, combustion of fuel droplets 
in engine, diagnosis of diseases, development of modern printing methods for inkjet printers, 
manufacture of new electronic and optical devices and a number of other 
areas~\cite{brutin2018, larson2014, erbil2012, Zang2019, Pauchard2018, Shao2020}.

The nonuniform mass flow and heat transfer during a drop evaporation have a crucial influence 
on the evaporation kinetics. In particular, they change the temperature distribution along 
the droplet surface and can cause thermocapillary convection inside the droplet due to the 
temperature-dependent surface tension. Axisymmetric Marangoni flows can often 
be observed at room temperature and under atmospheric 
pressure~\citep{Savino2004,Kang2004,HuLarsonReversal,ristenpart2007}.

A thermocapillary fluid flow with the broken axial symmetry can spontaneously appear inside 
a substantially heated droplet with an intense evaporation and sufficiently high fluid 
velocity. In a heated droplet, a Marangoni instability due to non-uniform temperature 
distribution can lead to the fluid motion of the particular form of longitudinal rolls or 
hydrothermal waves (HTWs), as was first demonstrated in~\cite{sefiane2008} and 
confirmed in numerous consequent 
studies~\cite{sefiane2010,brutin2011,sobac2012,carle2012,sefiane2013,karapetsas2012,saenz2014,zhu2019}. 
HTWs are time-periodic traveling waves in the liquid, induced by instabilities of the steady
Marangoni convection, which were originally predicted for the liquid layer on a rigid plate 
by Smith and Davis~\cite{smith1983,smith1986} and later observed 
in~\cite{schwabe1992,riley1998}. According to Smith and Davis~\cite{smith1983,smith1986}, the 
disturbance propagates perpendicular to the surface temperature gradient at small Prandtl 
numbers and almost exactly in the upstream direction at large Prandtl 
numbers~\cite{smith1983,smith1986}.
For a flat liquid layer, HTWs and BM
convection usually occur under qualitatively different conditions. 
While the Benard-Marangoni effect becomes more pronounced with an increase 
of the vertical temperature gradient in films, the appearance of HTWs
is associated with a horizontal temperature gradient
(for example, they can appear during lateral heating). In the case of evaporating 
sessile droplets, the competition between HTWs and BM convection is more complex.

The HTWs in droplets were originally identified by Sefiane et al. in 2008~\cite{sefiane2008}, 
when the evaporation of sessile droplets of ethanol, methanol, water, and Fluorinert FC-72
was investigated. Later more detailed observations of HTWs have been reported for the ethanol 
droplets on four different substrates with different thermal conductivities~\cite{sefiane2010}. 
HTWs moving in azimuthal direction have been observed for volatile droplets of ethanol and 
methanol. For FC-72, convective cells emerged near the droplet apex and moved towards the 
contact line. The number of waves has been found to increase with the increase of substrate 
thermal conductivity and its temperature, and to decrease with the decrease of the droplet 
height. On the other hand, no hydrothermal instability has been found in evaporating water 
droplets.

Three stages of the evaporation process with thermoconvective instabilities have been 
distinguished in~\cite{brutin2011} in the droplets of methanol, ethanol, and FC-72 on a 
polytetrafluoroethylene substrate: heating of the drop (which reaches almost the substrate 
temperature), evaporation with thermoconvective instabilities (the heat flux is maximal at 
this stage), and evaporation without thermoconvective patterns (the droplet is more like a 
film at this stage). The main phase is characterized by the presence of convection cells, and 
the origin of the cells' movement is associated with the temperature difference between the 
heated substrate and the ambient temperature. 
Temporal evolution of the HTWs was also described by means of simultaneous 
thermal and fluid motion
observations in a drying ethanol droplet on a heated substrate~\cite{sobac2012}.
HTWs traveling in azimuthal direction in 
a droplet of ethanol were demonstrated to take place both under normal gravity and 
microgravity conditions, so that the gravity effects on the behavior of HTWs are 
small~\cite{carle2012}.

Sefiane et al.~\cite{sefiane2013} demonstrated that HTWs in FC-72 droplets
are bulk waves that extend across the entire droplet volume.
Also, the influence of the HTWs on the heat transfer in the solid substrate
and on evaporation rate spatial distribution was studied in~\cite{sefiane2013}.
A specific kind of HTWs was determined by Karapetsas et al. 
in~\cite{karapetsas2012}, where three-dimensional spiral-like time-periodic traveling waves 
were observed that are organized radially and are uniform along the azimuthal angle. 
Shi et al.~\cite{shi2017} determined critical Marangoni 
numbers for the emergence of the Marangoni convection instabilities in a sessile 
droplet with low volatility on a heated substrate by a series of three-dimensional computer 
simulations.

New features of the problem in question have been revealed by Semenov et 
al.~\cite{semenov2017}, who have found that the cell patterns they determined in the droplet 
with a large Prandtl number represent the BM cell structure rather than HTWs. The authors 
have observed Marangoni instability in an ethanol droplet of capillary size on a heated and 
highly conductive substrate. They also performed numerical simulations by one-sided model 
described in~\cite{semenov2017PRE}, where they took into account, in particular, an unsteady 
diffusion-limited evaporation for non-isothermal pinned sessile droplet and the Stefan flow 
in the gas. After solving the heat transfer equation in the droplet bulk and carrying out the
numerical simulation, the results obtained have agreed well with the experiment for the
contact angle $\theta\approx 29.3^\circ$. At the first stage, a torus roll adjacent to the 
contact line appeared; later it is destabilized and split into several three-dimensional 
unsteady BM cells occupying the whole droplet. In the third stage, the BM cells drifted towards 
the contact line where the Marangoni forces are stronger, and the flower pattern was observed 
with about $15$ BM cells along the circumference, which corresponds to the spatial period 
about twice the local liquid thickness. The authors found that the number 
of BM cells increased with the decrease of the droplet height. 

The interplay of the BM convection cells and HTWs has been observed in the recent 
work~\cite{wang2019} in a volatile sessile droplet of silicone oil on a copper substrate at 
constant contact line mode over a wide range of substrate temperatures. When the substrate 
temperature was higher than room temperature, the BM quasistationary convection cell structure 
was observed and found to evolve into an oscillating state, as the evaporation progresses. 
When the substrate was colder than the surrounding air, the oscillating BM cells were observed 
only when the temperature of the substrate is higher than $10.6^{\circ}$C. The cell 
size became larger and their number was smaller when the temperature of the substrate goes up 
above the temperature of the surrounding gas. When the substrate temperature is above 
$14.3^{\circ}$C, the coexistence of advancing HTWs and BM 
vibrational convection was observed. The paper reports critical values of the contact angle at 
which the transition between different types of convection occured.

In particular, the authors~\cite{wang2019} considered the case when the substrate temperature
is $7.81$ K higher than the ambient temperature. Here, at the first stage, a flower structure 
of BM cells near the triple line emerged. With decreasing the contact angle, the cells grew in 
size and occupied the central region of the droplet surface (Fig.4 in~\cite{wang2019}). 

The qualitatively similar flower structures of BM cells near the triple line
at an intermediate stage of the droplet evaporation were observed in both 
publications~\cite{semenov2017,wang2019}. At the same time, the spatial 
structures of BM cells at sufficiently small contact angles,
which were observed in~\cite{wang2019} and numerically obtained in~\cite{semenov2017},
differ substantially.
The similarity of the results obtained 
in~\cite{semenov2017,wang2019} is expected as the droplet parameters 
considered in the two papers are close to each other. Also the relevant physical properties of 
$0.65$ cSt silicone oil and properties of the ethanol are similar, and highly conductive 
heated substrates at similar temperatures are employed in both works. It is therefore surprising 
that the results are quite different at a later stage of the droplet evaporation.
The purpose of this paper is to
analyze the behavior of unsteady BM cells in volatile sessile droplets on 
heated substrates including the stage of comparatively small contact angles. It is also of interest 
to identify the main physical mechanisms, which are responsible for the behavior studied and have 
to be taken into account in the corresponding simulations.

We have theoretically investigated the Marangoni instability patterns in evaporating sessile 
droplets of ethanol and silicone oil for the particular parameters which are close to each of 
the experiments~\cite{semenov2017,wang2019}. Based on the three-dimensional numerical 
simulations, the fluid flows and geometric characteristics of the BM patterns in the droplet
have been described. The mathematical model was developed that takes account 
of heat transfer in all three phases, hydrodynamics, vapor diffusion in air, Stefan flow
in the gas and thermal radiation. 
Using ANSYS Fluent software, three-dimensional computer simulations were carried out and
the unsteady temperature distributions and velocity fields in the drop were obtained. 
We have theoretically recovered both the flower structure of BM cells near triple line
for sufficiently large contact angles and the BM cell pattern where larger BM cells occupy the
fluid-gas interface for smaller contact angles.

The paper is organized as follows. Sec.~\ref{mathmodel} describes mathematical model
and method of numerical solution. Relevance of various physical effects 
for the problems under consideration are also analyzed in Sec.~\ref{mathmodel}.
In Sec.~\ref{ResultsSec} we present our main results.
Sec.~\ref{ConclusionSec} contains discussion.

\begin{table} 
\caption{The notations and parameter values used in the calculations.}
\begin{center} 
\small
\begin{tabular}{ccccc} 
\hline
notation & quantity & unit & ethanol & $0.65$ cSt silicone oil \\
\hline
$R$           & Contact line radius & mm  &  $2.95$  &  $2.09$   \\
$\theta$      & Contact angle & degrees &  $29.2^\circ, 20^\circ, 10^\circ$  & $28.39^\circ, 20.49^\circ, 17.18^\circ$    \\
$T_s$         & Substrate temperature & K   &  $307.05$  &  $302.69$   \\
$T_a$         & Ambient air temperature & K   &  $297.55$  & $294.88$    \\
$\rho$        & Density & kg/m$^3$   &  $772.24$  & $760$    \\
$c_p$         & Specific heat &  J/(kg$\cdot$K)  &  $2602.3$  & $2000$    \\
$k$           & Thermal conductivity & W/(m$\cdot$K)   &  $0.14$  & $0.1$    \\
$\kappa=k/(\rho c_p)$ & Thermal diffusivity & m$^2$/s & $6.97\cdot 10^{-8}$ & $6.58\cdot 10^{-8}$\\
$k_{\text{air}}$ & Thermal conductivity of air & W/(m$\cdot$K) & $0.026$ & $0.026$\\
$\eta$        & Dynamic viscosity & kg/(m$\cdot$s)   &  $1.095 \cdot 10^{-3}$  & $4.94 \cdot 10^{-4}$    \\
$\sigma$      & Surface tension & mN/m   &  $2.062 \cdot 10^{-2}$  &  $1.54 \cdot 10^{-2}$   \\
$\sigma_T$    & Temperature derivative of surface tension & N/(m$\cdot$K)   &  $-8.979 \cdot 10^{-5}$   &  $-8 \cdot 10^{-5}$   \\
$L$           & Latent heat of evaporation & J/kg   &  $918 600$  & $223 000$  \\
$D$           & Diffusion constant of vapor in air & m$^2$/s   &  $11.81 \cdot 10^{-6}$  &  $5.6 \cdot 10^{-6}$   \\
$u_s$         & Saturated vapor density & kg/m$^3$   &  $0.219664$ & $0.469145$ \\
$\varepsilon$ & Emissivity & \   &  $0.92$  &  $0.91$ \\
\hline
\end{tabular}
\end{center} 
\label{NotationsTable}
\end{table}

\section{Mathematical model}
\label{mathmodel}

We consider three-dimensional nonstationary problem for an axially symmetrical sessile droplet
on a heated solid substrate. The surface is taken as a spherical cap. 
This approximation is valid under the condition $\text{Bo} \ll 1$,
i.e., when the influence of gravitation on the droplet shape is small.
Here $\text{Bo} = {\rho g h_0 R}/{(2 \sigma \sin{\theta})}$
is the Bond number, and $h_0$ is the droplet height. We use the notations in Table~\ref{NotationsTable}.

\subsection{Diffusion of vapor in air}

The dynamics of the vapor concentration $u$ in the surrounding atmosphere
is described by the diffusion equation

\begin{equation}
\dfrac{\partial u}{\partial t} = D \Delta u.
\end{equation}

The boundary conditions are as follows: $u = u_s$ on the drop surface, $u = 0$ far away from the drop, 
$\partial u / \partial r = 0$ and $\partial u / \partial z = 0$ on the axes $r = 0$ and $z = 0$, respectively.

Semenov et. al derive in~\cite{semenov2017PRE} and employ in~\cite{semenov2017}
analytical equations for the unsteady 
diffusion-limited evaporation from sessile droplets, which agree with
the numerical results of~\cite{barash2009}.
However, these unsteady effects are small on time scales larger
than the time required for the Brownian particle to pass the characteristic length $R$.
i.e., for $t \gtrsim R^2/D \sim 0.7$ s., for the problems under consideration.
Therefore, the evaporation kinetics can be considered as a steady state process
and one can use the quasi-stationary approximation 
\begin{equation}
\Delta u = 0
\label{us}
\end{equation}
with the same boundary conditions together with the spherical cap approximation.

The latter problem is mathematically equivalent to the one
for the electrostatic potential of a charged conductor with a shape formed 
by two intersecting spheres. 
Such a problem has been solved analytically in~\cite{lebedev} in terms of toroidal coordinates.
Based on this circumstance, the analytic solution 
for Eq.~(\ref{us}) with the above boundary conditions, that describes 
the inhomogeneous evaporation flux density $J(r)$ from the surface of the 
evaporating droplet, was presented in~\cite{deegan2000} in the form
\begin{equation}
J(r) = |D \nabla u| = \dfrac{D u_s}{R} \bigg(\dfrac{\sin{\theta}}{2} + \sqrt{2} (x(r) + \cos{\theta})^{3/2}\displaystyle\int_{0}^{\infty} \dfrac{\cosh{(\theta\tau)}}{\cosh{(\pi \tau)}} \tau \tanh((\pi - \theta)\tau) P_{-1/2+i\tau} (x(r)) d\tau\bigg),\label{analytical}
\end{equation}
where 
\begin{equation}
x(r) = \bigg(r^2 \cos{\theta}/R^2 + \sqrt{1-r^2 \sin^2{\theta} / R^2}\bigg) \bigg\slash (1-r^2/R^2)
\end{equation}
and $P_{-1/2+i\tau} (x)$ is the Legendre polynomial.
Expr.~(\ref{analytical}) can be conveniently 
approximated as~\citep{deegan2000,hu2002}
\begin{equation}\label{Jr}
J (r) = J_0(\theta) \bigg(1-\dfrac{r^2}{R^2}\bigg)^{-\lambda(\theta)},
\end{equation}
where $\lambda(\theta) = 1/2 - \theta/\pi$.
The coefficient $J_0 (\theta)$ in (\ref{Jr}) can be fitted as follows:

\begin{equation}
\dfrac{J_0 (\theta)}{1-\Lambda(\theta)} = J_0(\pi/2) (0.27 \theta^2 + 1.3),
\end{equation}

\begin{equation}
\Lambda(\theta) = 0.22(\theta-\pi/4)^2+0.36,
\end{equation}

\begin{equation}\label{J_0}
J_0(\pi/2) = \dfrac{D u_s}{R},
\end{equation}
where $u_s$ is the saturated vapor density at the droplet free surface. 
It is numerically obtained in~\cite{hu2002} that Eqs.(\ref{Jr})--(\ref{J_0}) represent an accurate
fit for the solution of Eq.(\ref{us}) with the above boundary conditions,
within the spherical cap approximation.
The fit is valid in a wide range of contact angles, and its accuracy
does not depend on the parameters such as $D$ and $R$. For this reason, we employ 
the Eqs.(\ref{Jr})--(\ref{J_0}) in the simulations.

Saturated vapor pressure 
is determined by the Antoine equation (with $p_s$ in (Pa) and $T$ in ($^\circ$K)):
\begin{equation}
\log_{10}{p_s\left(T\right)=A-\frac{B}{C+T}},  
\end{equation}
where $A$, $B$, and $C$ are empirical constants. 
We use $A=10.247$, $B=1599.039$, $C=-46.391$ for ethanol~\cite{Ambrose1970}.
For $0.65$ cSt silicone oil, we use $A=9.418$, $B=1509$, $C=-31.55$~\cite{saenz2014}.

The concentration of vapor $u_s$ at the drop surface is calculated from ideal gas law and defined by

\begin{equation}
u_s\left(T\right)=\dfrac{M p_s\left(T\right)}{\bar{R}T},
\label{u_s_antoine}
\end{equation}
where $\bar{R}$ is the universal gas constant and $M$ is molar mass of the vapor.

It follows from~(\ref{Jr}) that the mass loss resulting from vaporization 
occurs nonuniformly along the free surface of the liquid layer,
significantly increasing near the pinned contact line.
A nonuniform mass flow during evaporation and the corresponding heat transfer
change the temperature distribution along the droplet surface 
and cause Marangoni forces due to temperature-dependent surface tension.
These forces produce thermocapillary convection inside a droplet.

\subsection{Stefan flow in the gas}

Because the saturated vapor concentration is not much smaller 
than the density of air (see Table~\ref{NotationsTable}), it is natural to consider
the contribution of the Stefan flow in the gas
for the problems under consideration~\cite{fuchs1959}.

As a result of the convective mass transport in the gas, 
the evaporation rate increases as~\cite{semenov2017PRE}:
\begin{equation}
\tilde J(r)=\frac{u_g}{u_s} \ln\left(\frac{u_g}{u_g-u_s}\right) J(r),
\end{equation}
where $u_g=M_g p_g/(\bar{R}T)$, $p_g$ is the air pressure,
$M_g=M_{air}(1-X_v)+M_vX_v$ is the molar mass of gas near droplet free surface,
$M_{air}$ and $M_v$ are molar masses of corresponding air and vapor,
$X_v=u_s \bar{R} T/(M_v p_g)$ is the vapor molar fraction. 
Here, the surface temperature $T$ is a function of $r$.

It follows that the Stefan flow in the gas increases the evaporation rate 
by approximately $10\%$ and $18\%$ for droplets of ethanol and $0.65$ cSt silicone oil,
respectively. 

The results below in Sec.~\ref{ResultsSec} do not account for the Stefan flow in the gas.
However, our preliminary numerical calculations for the $0.65$ cSt silicone oil droplet
with parameters specified in Table~\ref{NotationsTable} and contact angle $20.49^\circ$
show that the influence of the Stefan flow on the resulting surface temperature distribution
is quite small.

\subsection{Hydrodynamics}

The governing equations for the fluid dynamics inside the drop are the Navier-Stokes equations 
and the continuity equation for the incompressible fluid

\begin{equation}\label{n-s}
\dfrac{\partial \text{\textbf{v}}}{\partial t} + (\text{\textbf{v}} \cdot \nabla) \text{\textbf{v}} + \dfrac{1}{\rho} \text{ grad } p = \nu \Delta \text{\textbf{v},}
\end{equation}

\begin{equation}\label{cont}
\text{div \textbf{v}} = 0,
\end{equation}

Here $\nu = \eta / \rho$ is kinematic viscosity, $\textbf{v}$ is fluid velocity, $p$ is pressure.

The boundary conditions for the equations (\ref{n-s}-\ref{cont})
include the no-slip condition $\textbf{v} = 0$ at the substrate--fluid interface and
the condition at fluid--gas interface which involves the Maranogoni forces
associated with the temperature dependence of the surface tension.
In general, the latter boundary condition can be expressed as~\cite{LL6}
\begin{equation}
\left(p-p_v-\sigma\left(\frac1{R_1}+\frac1{R_2}\right)\right)n_i=
\eta\left(\frac{\p v_i}{\p x_k}+\frac{\p v_k}{\p x_i}\right)n_k-
\frac{\p\sigma}{\p x_i},
\end{equation}
where the unit vector $\mathbf{n}$ is directed towards the vapor
along the normal to the surface.
Taking the tangential component of this equation,
one finds
\begin{equation}
\frac{d\sigma}{d\tau}=
\eta\left(\frac{\p v_\tau}{\p n}+\frac{\p v_n}{\p\tau}-v_\tau\frac{d\varphi}{d\tau}\right).
\label{vtau-boundary}
\end{equation}
Here $d\sigma/d\tau = \sigma_T\cdot dT/d\tau$, 
$\tau_i$ are the components of the unit vector $\pmb{\tau}$
tangential to the surface, and
$\varphi$ is the angle between the normal vector to the drop surface
and the vertical axis.
We note that the ANSYS Fluent simulation package allows assigning directly
the Marangoni stress at the droplet free surface,
so Eq.~(\ref{vtau-boundary}) is already implemented in the software 
and a user does not need to employ it.

\subsection{Heat transfer in the drop}

The temperature distribution is obtained with the heat
conduction equation

\begin{equation}\label{thermal}
\dfrac{\partial T}{\partial t} + \textbf{ v} \cdot \nabla T = \kappa \Delta T.
\end{equation}

The heat conduction inside the substrate is obtained with

\begin{equation}
\dfrac{\partial T}{\partial t}  = \kappa \Delta T.
\label{thermal_sub}
\end{equation}

The boundary conditions for the thermal conduction include
the isothermal boundary condition at the lower bound of substrate $T=T_s$, 
the adiabatic boundary condition at the substrate-gas interface $\partial T_s/\partial n=0$,
and the continuity of the heat flux at the substrate-fluid interface
$k_s\partial T_s/\partial z=k\partial T_L/\partial z$. 
Here $T_L$ and $T_S$ correspond to the liquid and substrate temperature, $k$ and $k_S$  are 
thermal conductivities of liquid and substrate, respectively.

Following the experiments~\cite{semenov2017,wang2019}, the calculations 
in the present work involve a highly conducting substrate.
In this case, the boundary conditions can be written simply as $T=T_s$ at the substrate-fluid
interface.

The boundary condition for the heat transfer at the droplet surface is 
\begin{equation}
\dfrac{\partial T}{\partial n} = - \dfrac{Q_0(r)}{k}.
\end{equation}
Here, 
\begin{equation}
Q_0(r) = -k_{\text{air}} \bigg(\dfrac{\partial T}{\partial n}\bigg)_{\text{air}} + L\tilde{J}(r) + \dot{q}_r
\label{Q_0}
\end{equation}
is the total heat flux from the droplet surface, which includes
latent heat of evaporation, effects of the thermal conduction in liquid and gas and thermal radiation.
The estimates below show that all these effects are important for droplets evaporating 
from heated substrates. 
Also, \textbf{n} is a normal vector 
to the droplet upper surface, $k_{air}\big(\partial T/\partial n\big)_{air}$ is the heat flow 
associated with non-uniform temperature distribution in an ambient gas, $k_{air}$ is the thermal 
conductivity of air, $J\left(r\right)$ is the local evaporation rate determined by (\ref{Jr}),
\begin{equation}
\dot{q}_r = \sigma_B \varepsilon (T^4 - T_a^4)
\label{Eq:radiation}
\end{equation}
is the net radiation flux between the 
drop and the environment; here $\sigma_B$ is the Stefan-Boltzman constant, $\varepsilon$ is the 
emissivity; $T$ and $T_a$ are the droplet surface temperature and temperature of the ambient gas, respectively.

The relation~(\ref{Eq:radiation}) can be employed under the following condition.
First, it is important that the droplet size is much larger than the characteristic wavelengths
of the thermal radiation, which are determined by the Planck distribution.
Under this condition the emissivity almost does not depend on the frequency~\cite{kattawar1970}.
And vise versa, for objects with dimensions much smaller than the thermal wavelength,
super-Planckian far-field radiative heat transfer is possible~\cite{thompson2018,cuevas2018}.
Secondly, it is important that the temperature difference inside the drop is much smaller than
the temperature difference between the droplet and the ambient gas.

Let's obtain the relation for $\left({\partial T}/{\partial n}\right)_{\text{air}}$.

The quasi-stationary temperature distribution in the gas phase is
determined by the equation
\begin{equation}
\Delta T = 0
\end{equation}
with the boundary conditions: $T = T_s$ at the droplet surface (under the assumption that 
the temperature difference inside the drop is much smaller than
the temperature difference between the droplet and the ambient gas),
$T = T_a$ away from the droplet.
This problem is mathematically equivalent to Eq.~(\ref{us}) with the corresponding
boundary conditions, therefore,
one can employ the same formulas for the analytical solution.
It follows from~(\ref{Jr})--(\ref{J_0}) that
the quantity $J(r)/(Du_s)$ depends only on the geometry of the problem.
It follows from the mathematical equivalence that
\begin{equation}
\frac{J(r)}{Du_s} = \frac{1}{T_s-T_a}\left(\frac{\partial T}{\partial n}\right)_{\text{air}}.
\label{Eq:HeatFluxAnalogy}
\end{equation}

Therefore, the main boundary condition at the droplet surface takes the form~(\ref{Q_0}),
where $\left({\partial T}/{\partial n}\right)_{\text{air}}$ is calculated similarly to
Eqns.(\ref{Jr})--(\ref{J_0}):

\begin{eqnarray}
\bigg(\dfrac{\partial T}{\partial n}\bigg)_{\text{air}} &=& J_1(\theta) \bigg(1-\dfrac{r^2}{R^2}\bigg)^{-\lambda(\theta)},\\
\dfrac{J_1 (\theta)}{1-\Lambda(\theta)} &=& J_1(\pi/2) (0.27 \theta^2 + 1.3),\\
J_1(\pi/2) &=& \dfrac{T-T_a}{R}.
\end{eqnarray}

\subsection{Heat transfer in the gas phase}

Here, we estimate the relative strength of the heat fluxes
$LJ(r)$ and $k_{\text{air}} \left({\partial T}/{\partial n}\right)_{\text{air}}$,
which are summands in~(\ref{Q_0}).
If follows from~(\ref{Eq:HeatFluxAnalogy}) that the heat fluxes are equal when
\begin{equation}
T_s - T_a = \dfrac{LD u_s}{k_{\text{air}}}.
\label{Eq:DeltaT1}
\end{equation}

Hence, the contribution of the effect of heat conduction in air
matches the evaporation cooling heat flux for
\begin{equation}
T_s - T_a \approx
\begin{cases}
60.91\, K \quad \mbox{for ethanol,} \\
0.94\, K \quad \mbox{for hexanol,} \\
22.53\, K \quad \mbox{for silicone oil}
\end{cases}
\end{equation}
The present work considers the case $T_s-T_a \sim 10$ K for droplets of ethanol 
and silicone oil, so taking into account the effect of heat transfer in the gas phase
is worthwhile for obtaining quantitative results.

\subsection{Thermal radiation}

Here, we estimate the relative strength of the evaporative cooling heat flux
$LJ(r)$ and the thermal radiation heat flux~(\ref{Eq:radiation}) between the droplet
and the environment.

It follows from Eqns.(\ref{Jr})--(\ref{J_0}) that the heat fluxes are roughly equal for
\begin{equation}
T \sim \left(\frac{LD u_s}{\sigma_B\varepsilon R}+T_a^4\right)^{1/4}.
\end{equation}

Hence, the contribution of the effect of thermal radiation 
matches the evaporative cooling heat flux for
\begin{equation}
T_s - T_a \approx
\begin{cases}
70.6\, K \quad \mbox{for ethanol,} \\
1.61\, K \quad \mbox{for hexanol,} \\
30.69\, K \quad \mbox{for silicone oil}
\end{cases}
\end{equation}
The present work considers the case $T_s-T_a \sim 10$ K for droplets of ethanol 
and silicone oil, so taking into account the effect of thermal radiation
is worthwhile for obtaining quantitative results.

\subsection{Numerical simulation}

We solve the thermal conduction equation and the Navier-Stokes 
equations by the finite element method using 
computational fluid dynamics simulation package
ANSYS Fluent. The simulation includes several steps.
\begin{itemize}
\item \textit{Geometry and mesh generation}

We create three-dimensional geometry of the axially symmetrical droplet. The surface curvature radius and the 
droplet height are equal to $R/\sin{\theta}$ and $R(1/\sin{\theta}- \cot{\theta})$, respectively. The next step 
is to divide the simulation region into small computational cells. The number of cells is between 200 and 500 
thousand depending on contact angle.

\item \textit{Defining model and setting properties}

We choose the pressure-based solver type, transient mode for time, viscous (laminar) flow and enable the calculation 
of energy in the model. We specify all physical properties of the fluid such as density, viscosity, the temperature 
derivative of the surface tension $\sigma_T$, which allows us to specify the properties of Marangoni stress, etc.

\item \textit{Setting boundary conditions and solution method}

The boundary conditions are set according to Sec.~\ref{mathmodel}. The rate of heat loss~(\ref{Q_0})
is specified with 
a user-defined function (UDF) written in the C programming language. The solution algorithm 
is SIMPLE (Semi-Implicit Method for Pressure Linked Equations)~\cite{patankar}.
The main feature of this method is using a relationship between velocity and pressure corrections to 
enforce mass conservation and to obtain the pressure field. The algorithm is written in such a way that 
the continuity equation is automatically satisfied.

\item \textit{Post-processing}

We display the simulation results: velocity field, absolute values of velocity, and temperature distribution 
and then analyze the vortex structure and wave patterns.
\end{itemize}

\section{Results and discussion}
\label{ResultsSec}

\begin{figure}[ht]
\begin{center}
\begin{minipage}[h]{0.47\linewidth}
\includegraphics[width=0.75\textwidth]{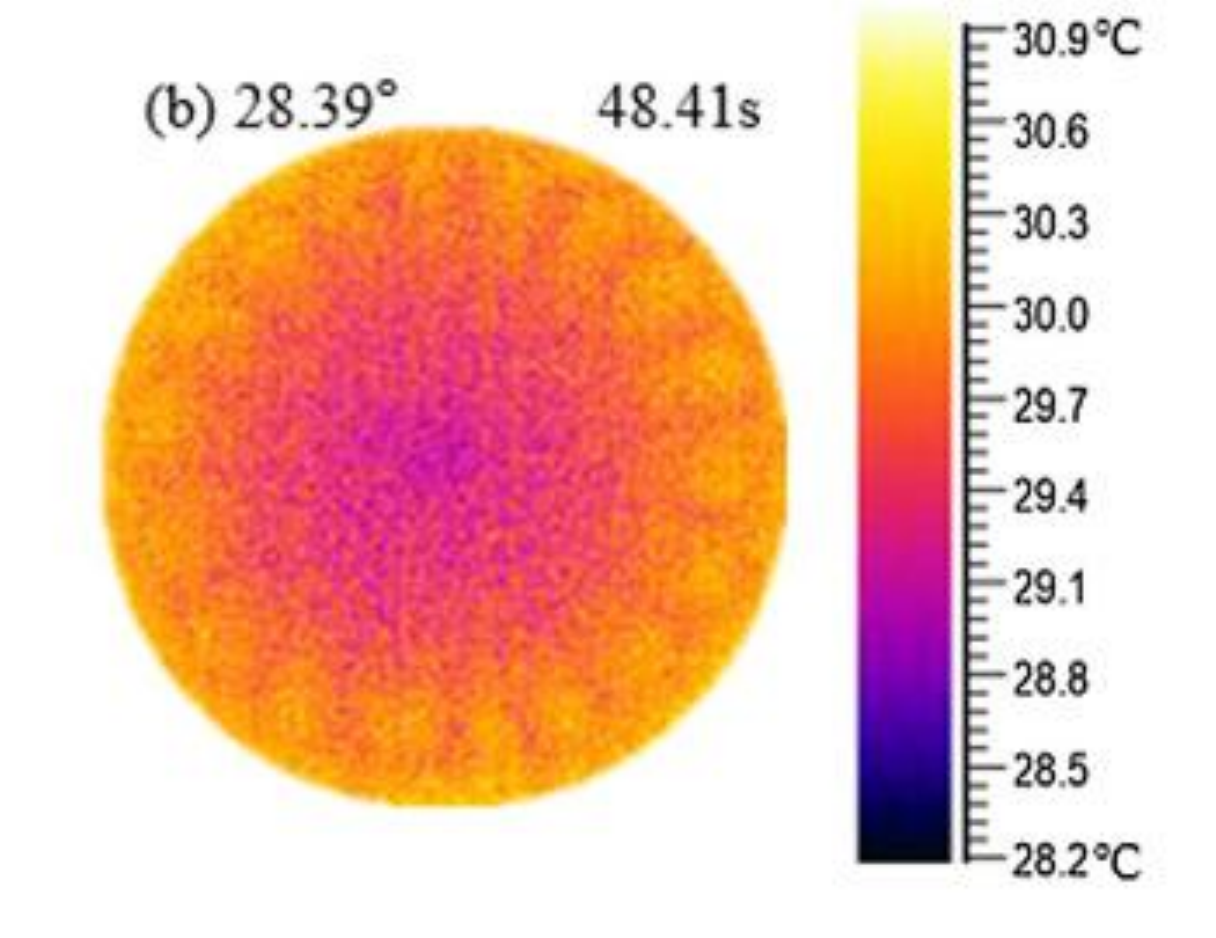}
\end{minipage}
\begin{minipage}[h]{0.47\linewidth}
\includegraphics[width=0.6\textwidth]{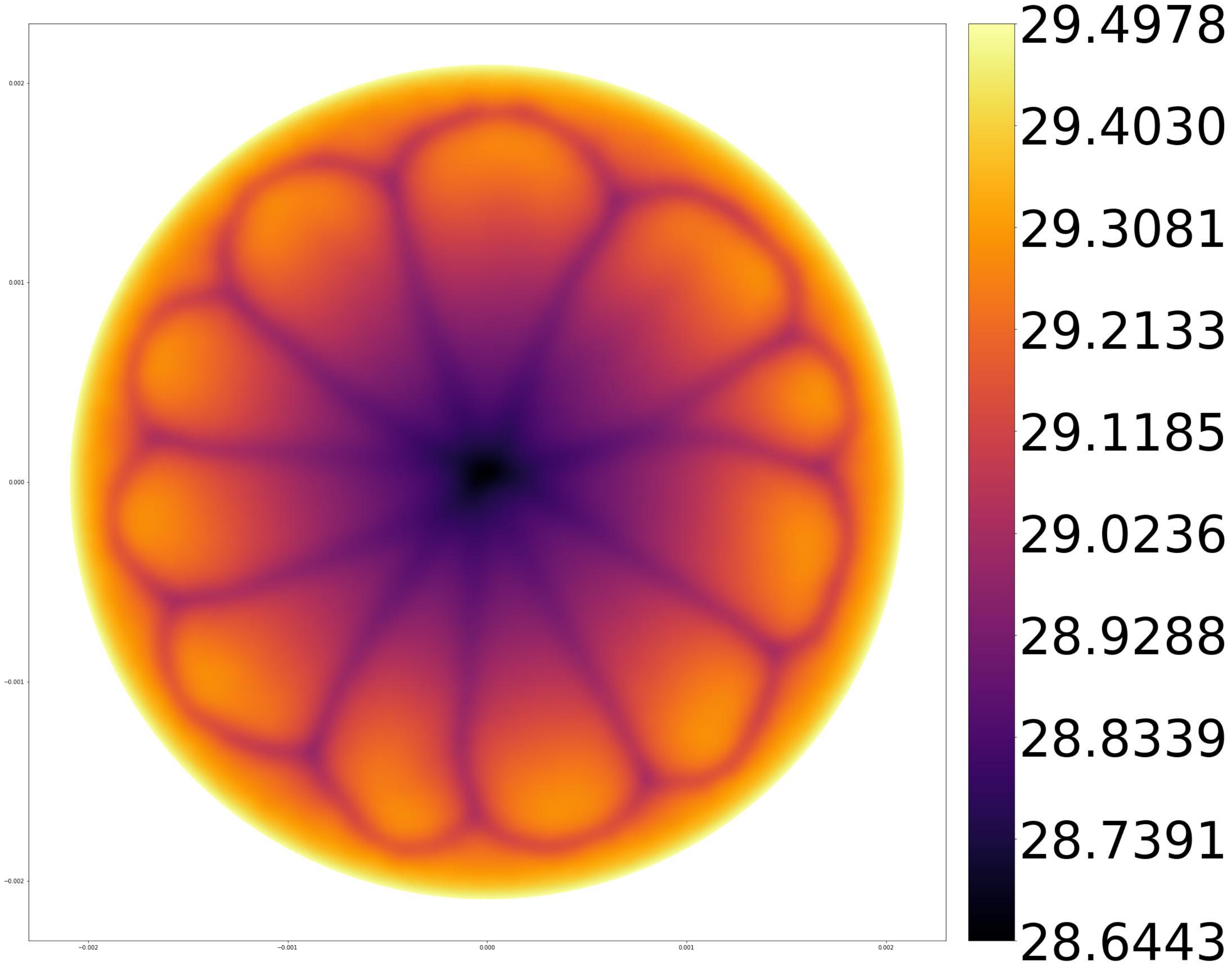}
\end{minipage}
\end{center}
\begin{center}
\begin{minipage}[h]{0.47\linewidth}
\includegraphics[width=0.75\textwidth]{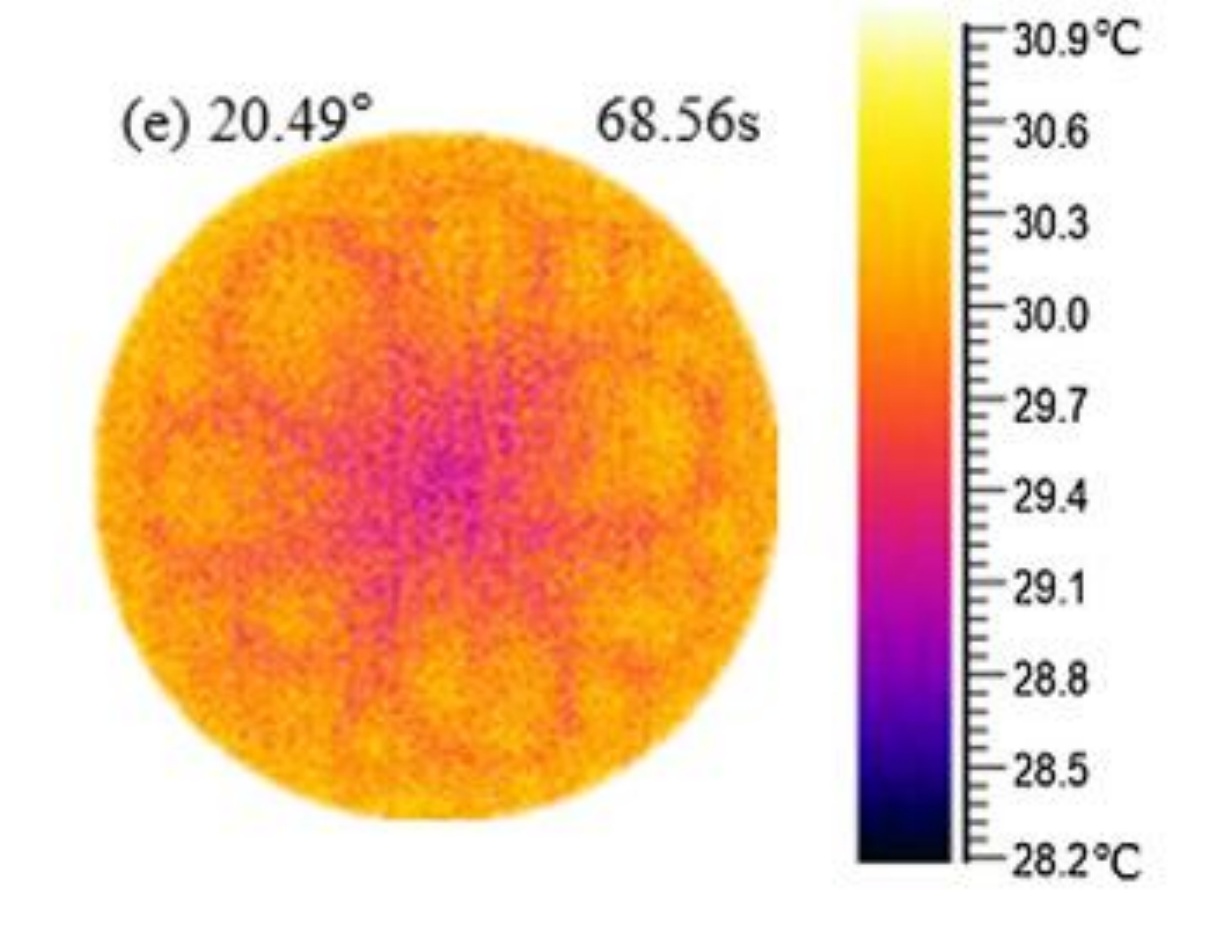}
\end{minipage}
\begin{minipage}[h]{0.47\linewidth}
\includegraphics[width=0.6\textwidth]{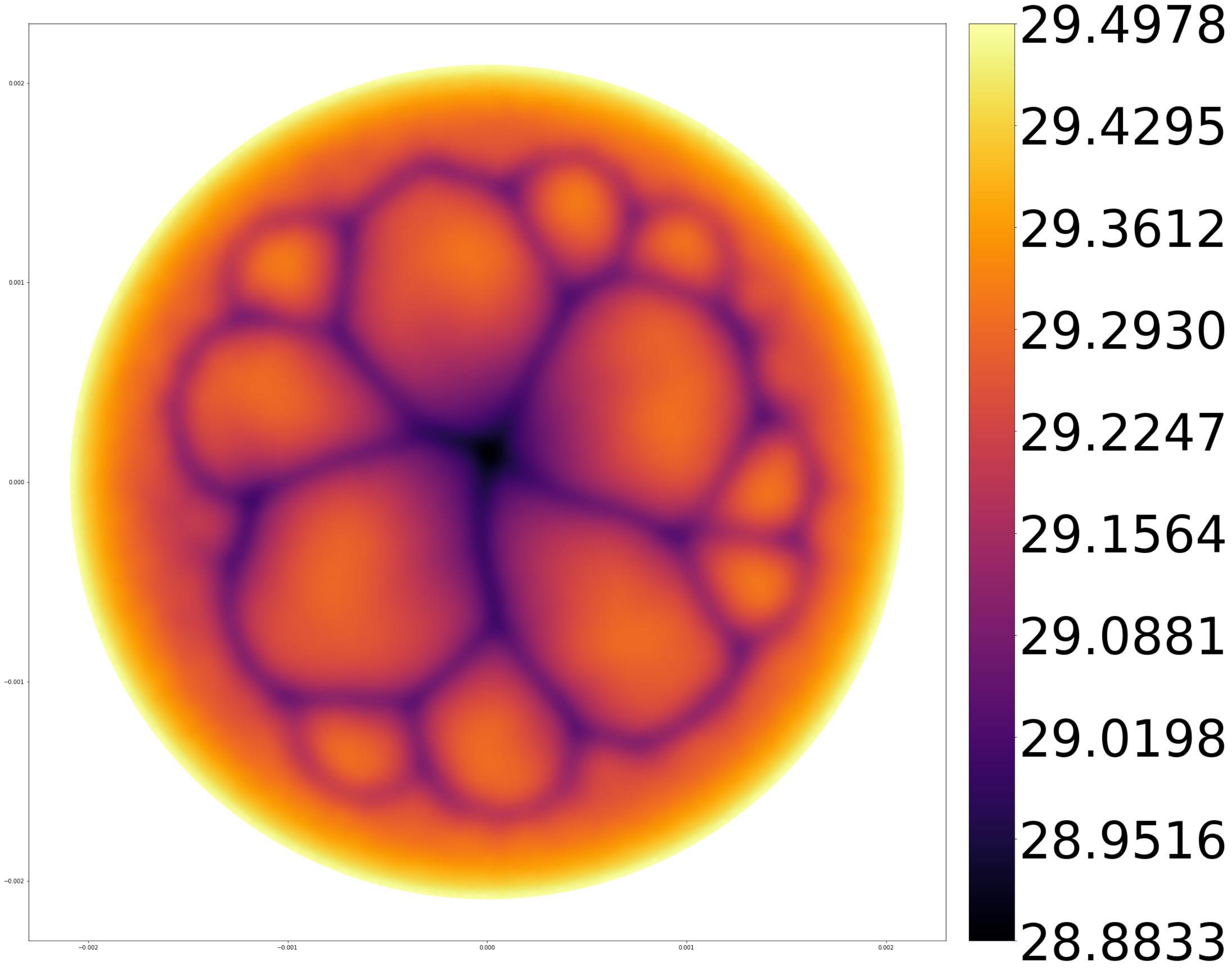}
\end{minipage}
\end{center}
\begin{center}
\begin{minipage}[h]{0.47\linewidth}
\includegraphics[width=0.75\textwidth]{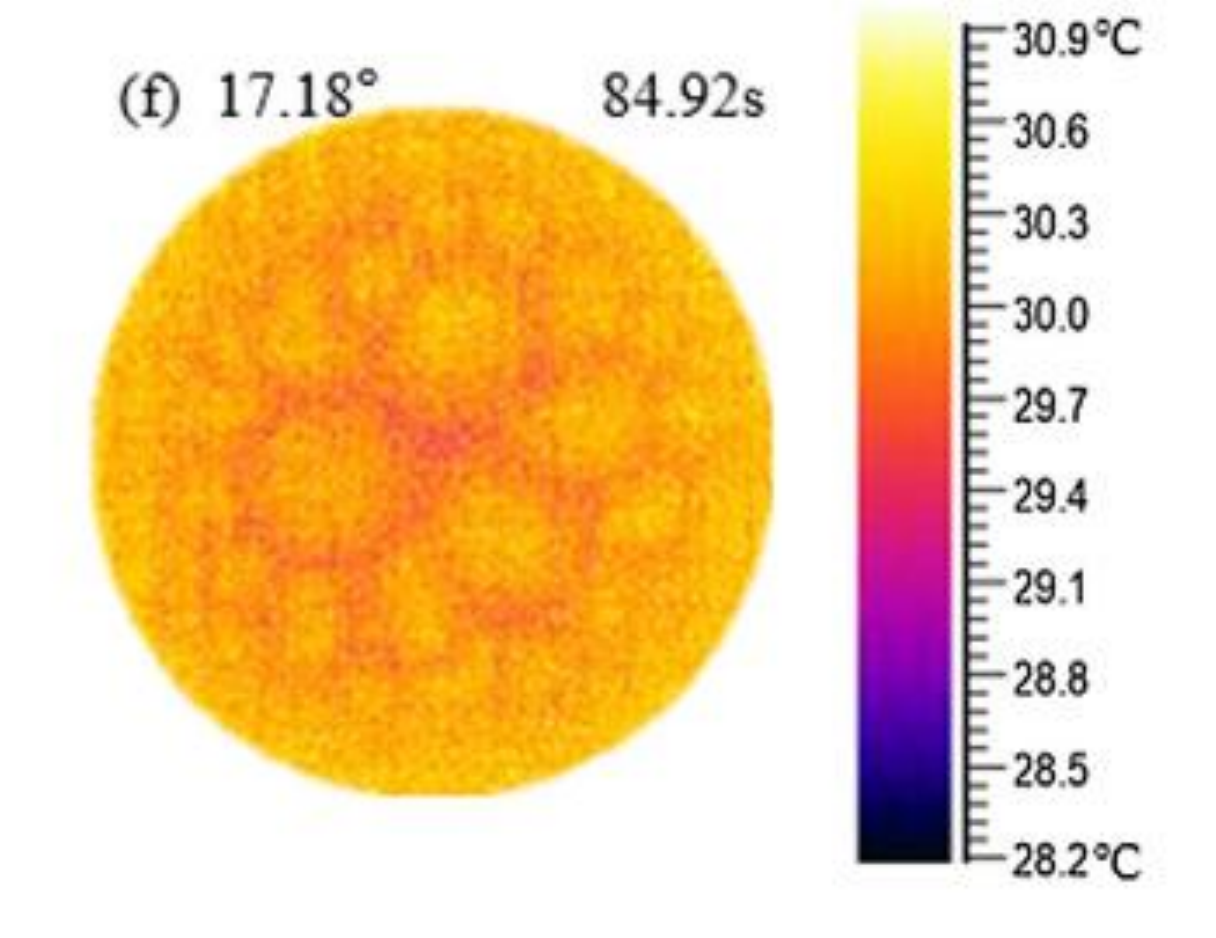}
\end{minipage}
\begin{minipage}[h]{0.47\linewidth}
\includegraphics[width=0.6\textwidth]{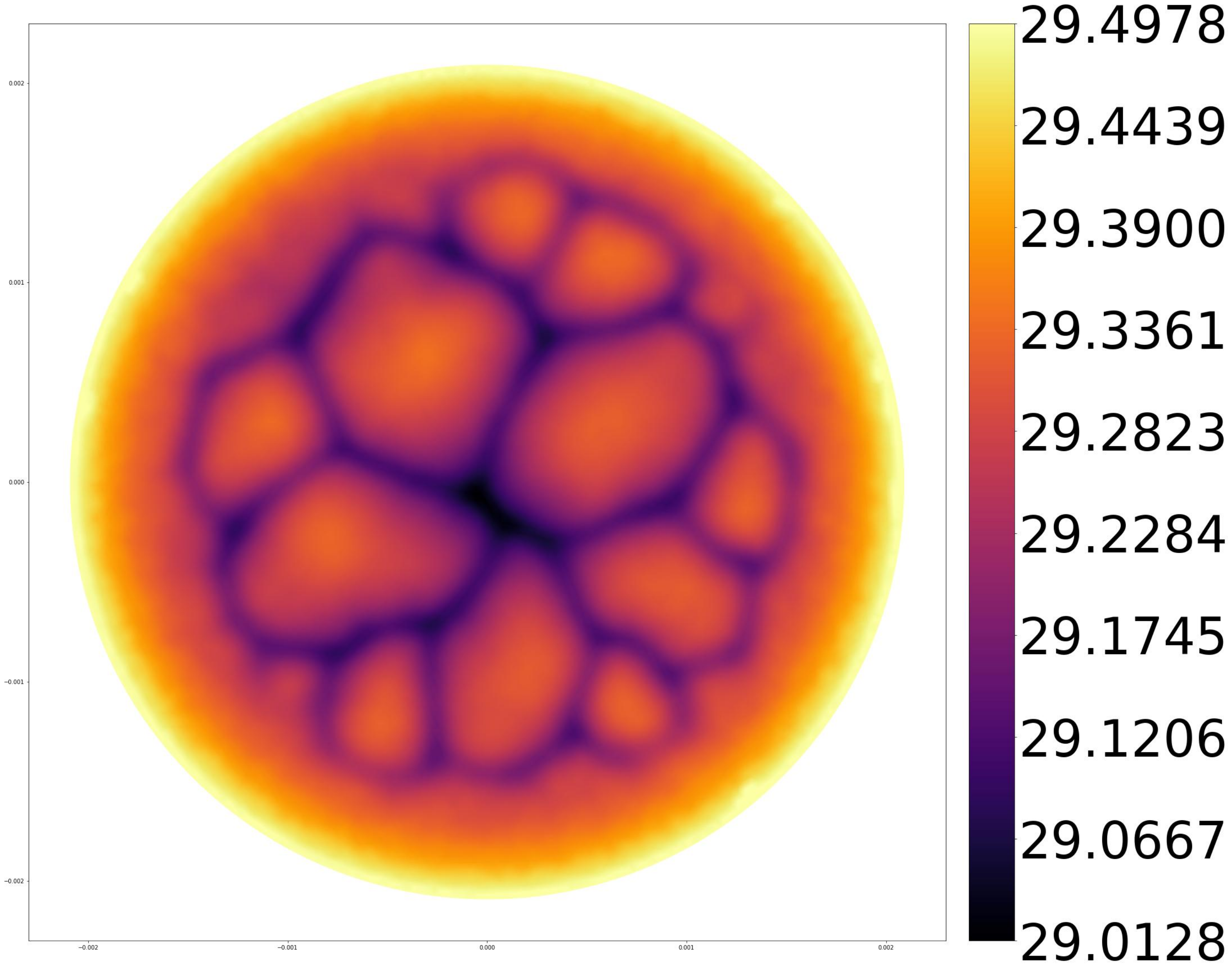}
\end{minipage}
\end{center}
\caption{Evolution of surface temperature patterns of the silicone
oil droplet. Left panel shows experimental data and is reprinted 
from Ref.~\cite{wang2019} (Copyright (2019), 
with permission from Elsevier).  Right panel 
shows the corresponding surface temperature patterns obtained by means of 
computer simulation
using the parameters from Table~\ref{NotationsTable}.}
\label{wang2019fig}
\end{figure}

\begin{figure}
\begin{center}
\includegraphics[width=0.24\textwidth]{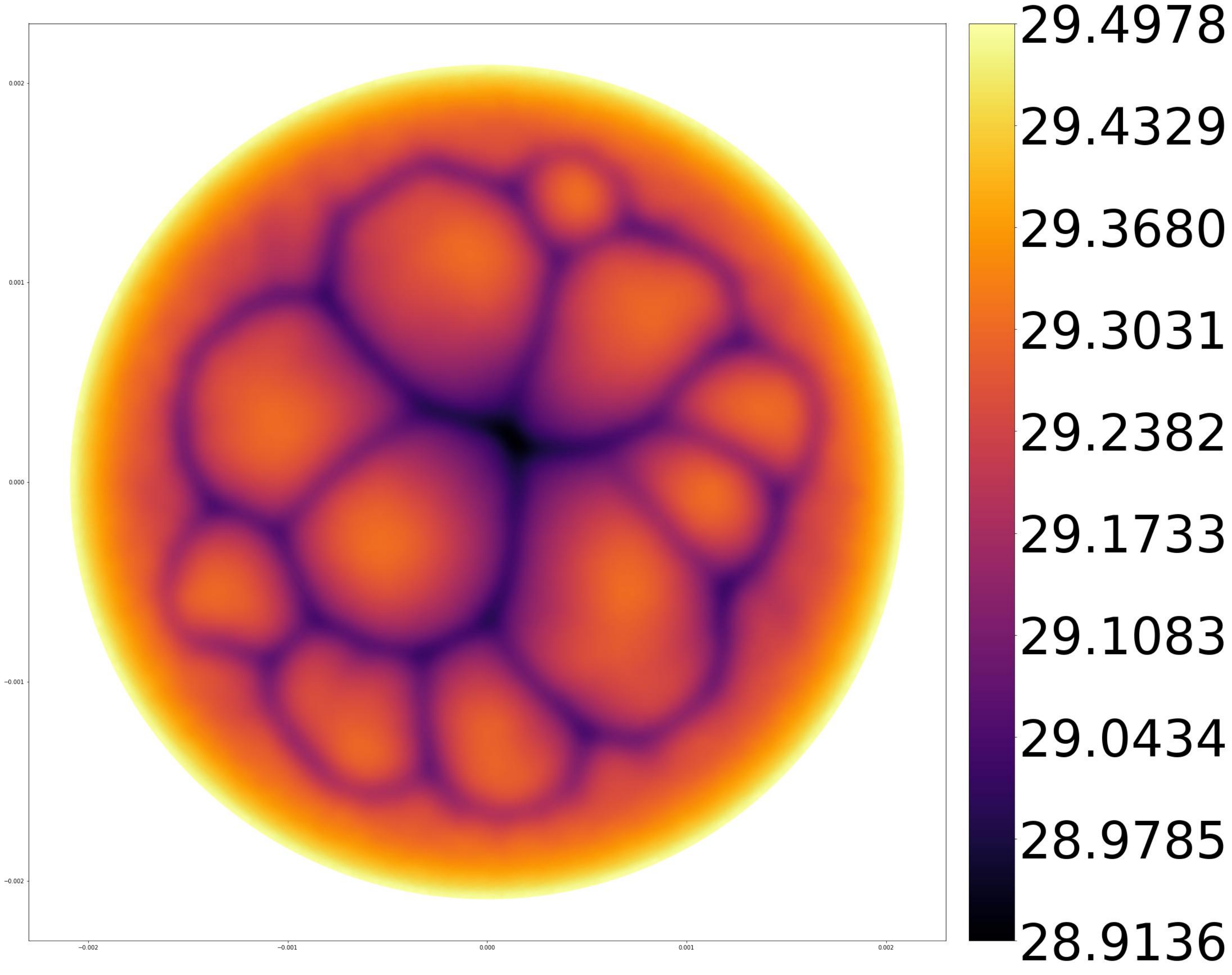}
\includegraphics[width=0.24\textwidth]{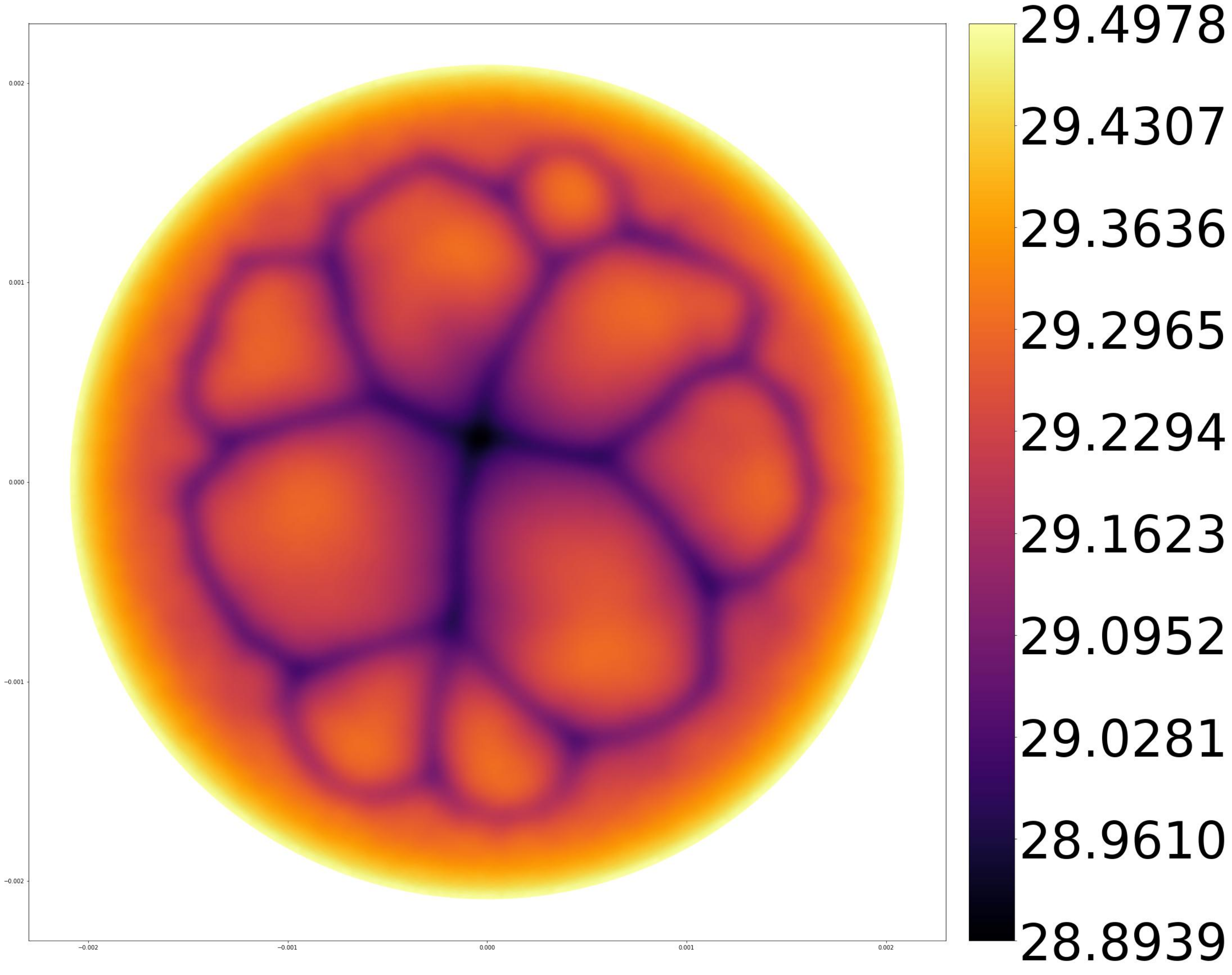}
\includegraphics[width=0.24\textwidth]{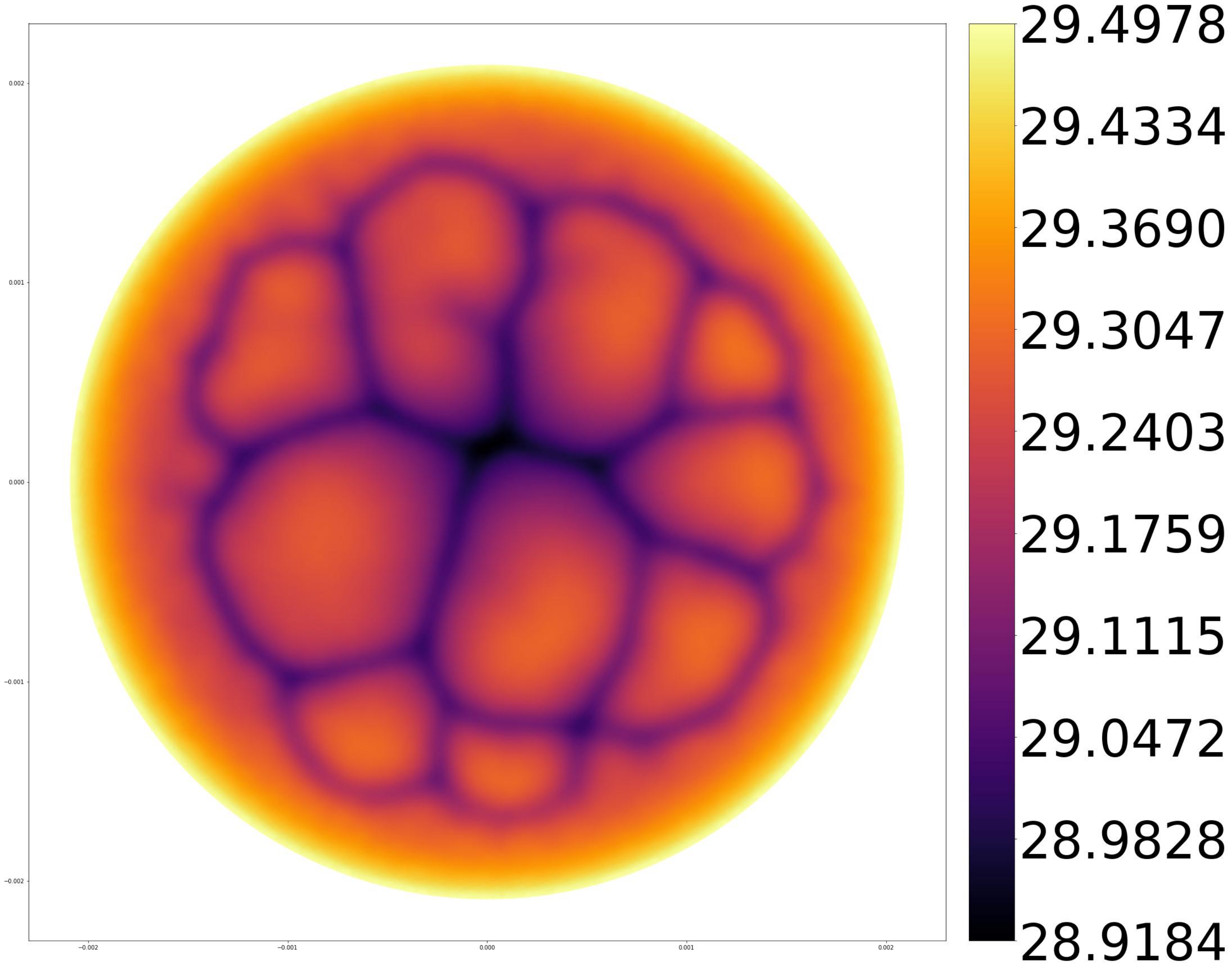}
\includegraphics[width=0.24\textwidth]{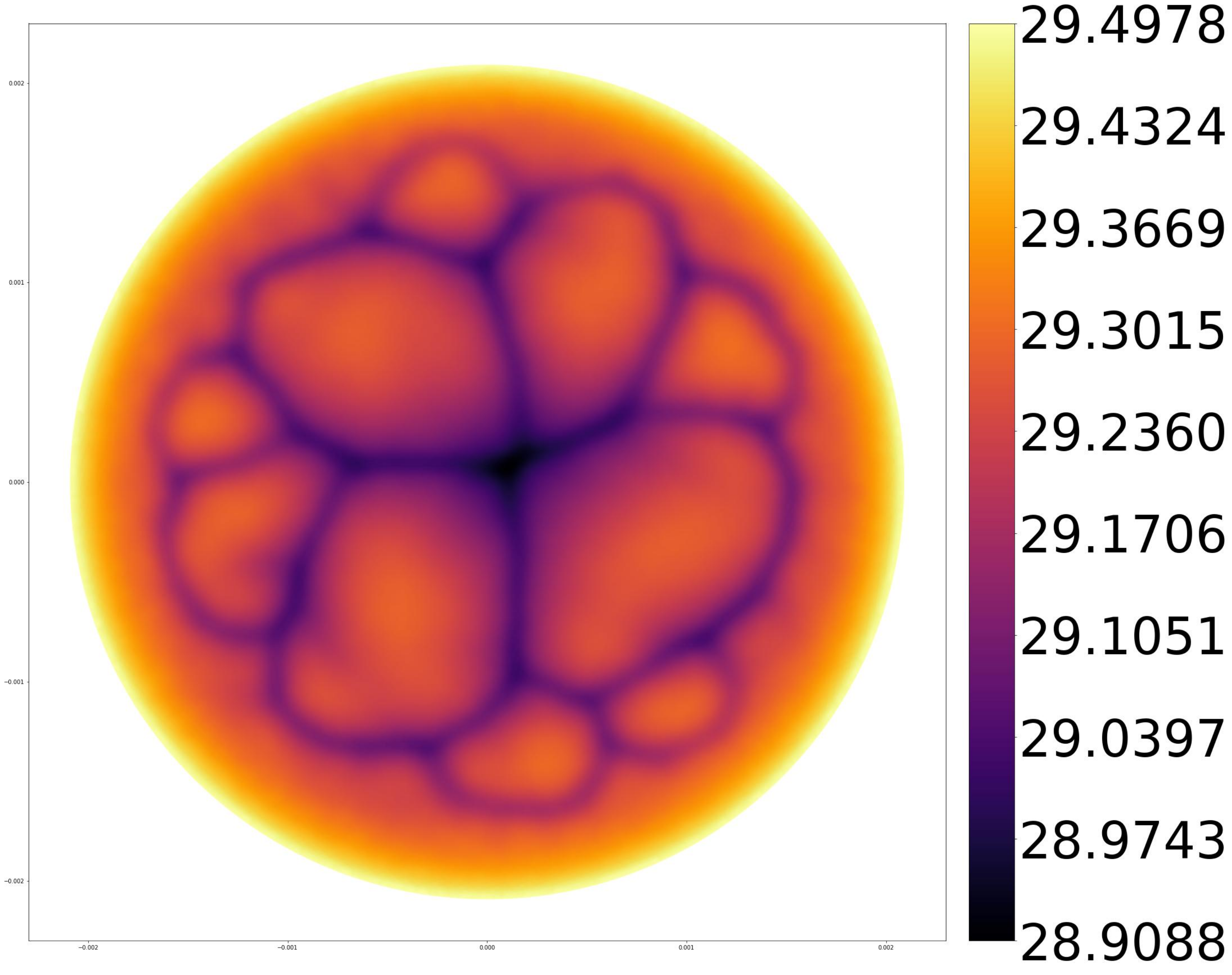}
\includegraphics[width=0.24\textwidth]{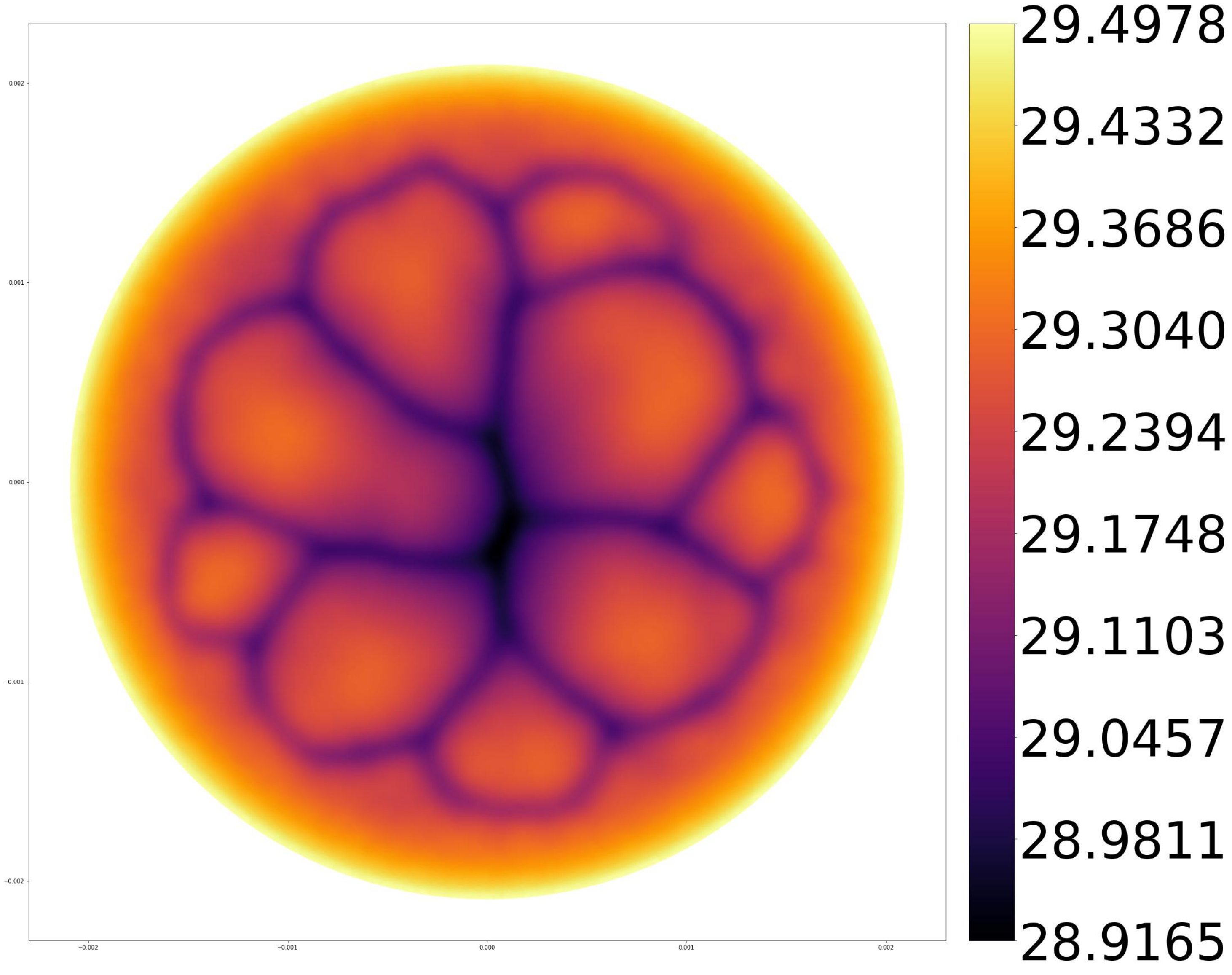}
\includegraphics[width=0.24\textwidth]{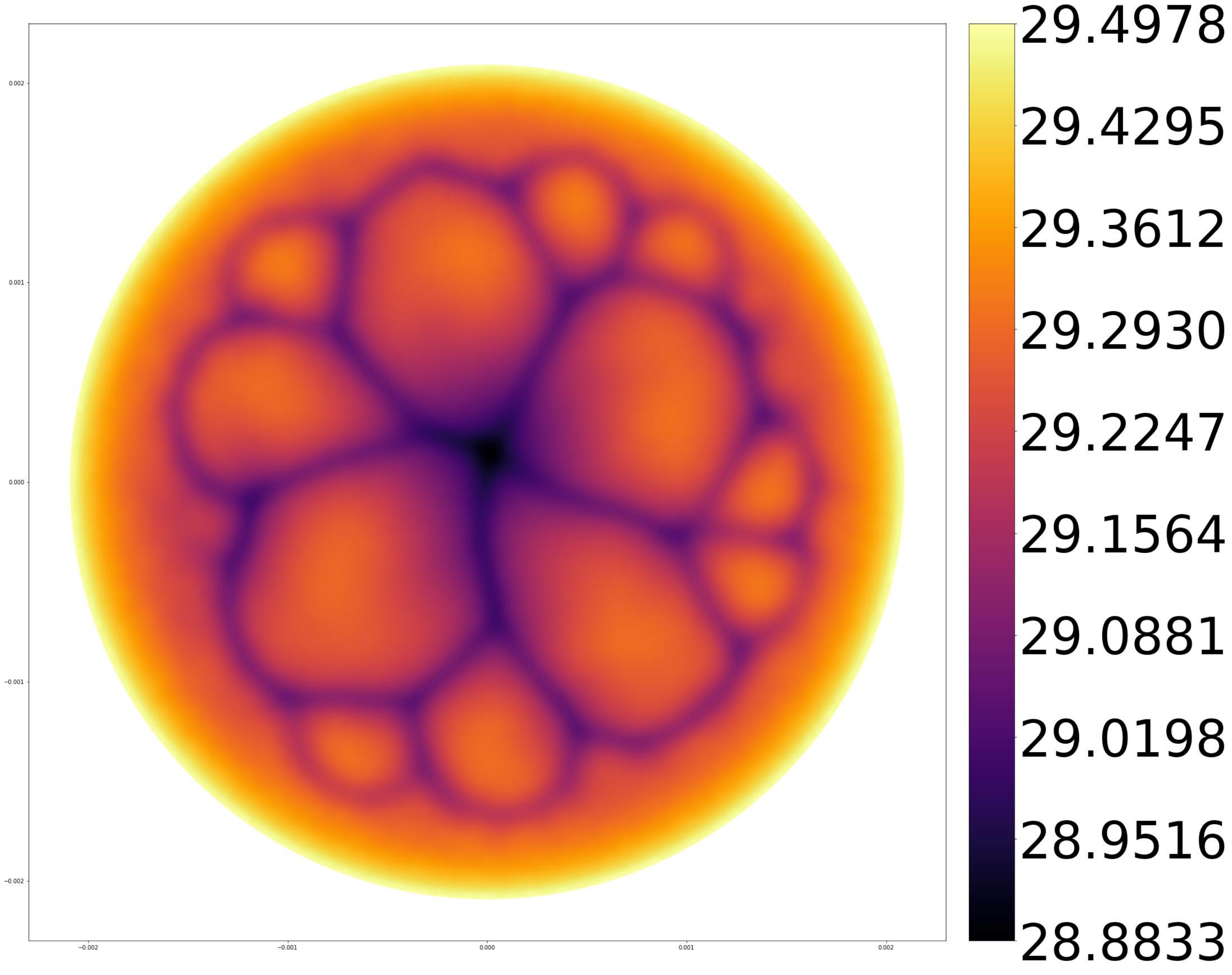}
\includegraphics[width=0.24\textwidth]{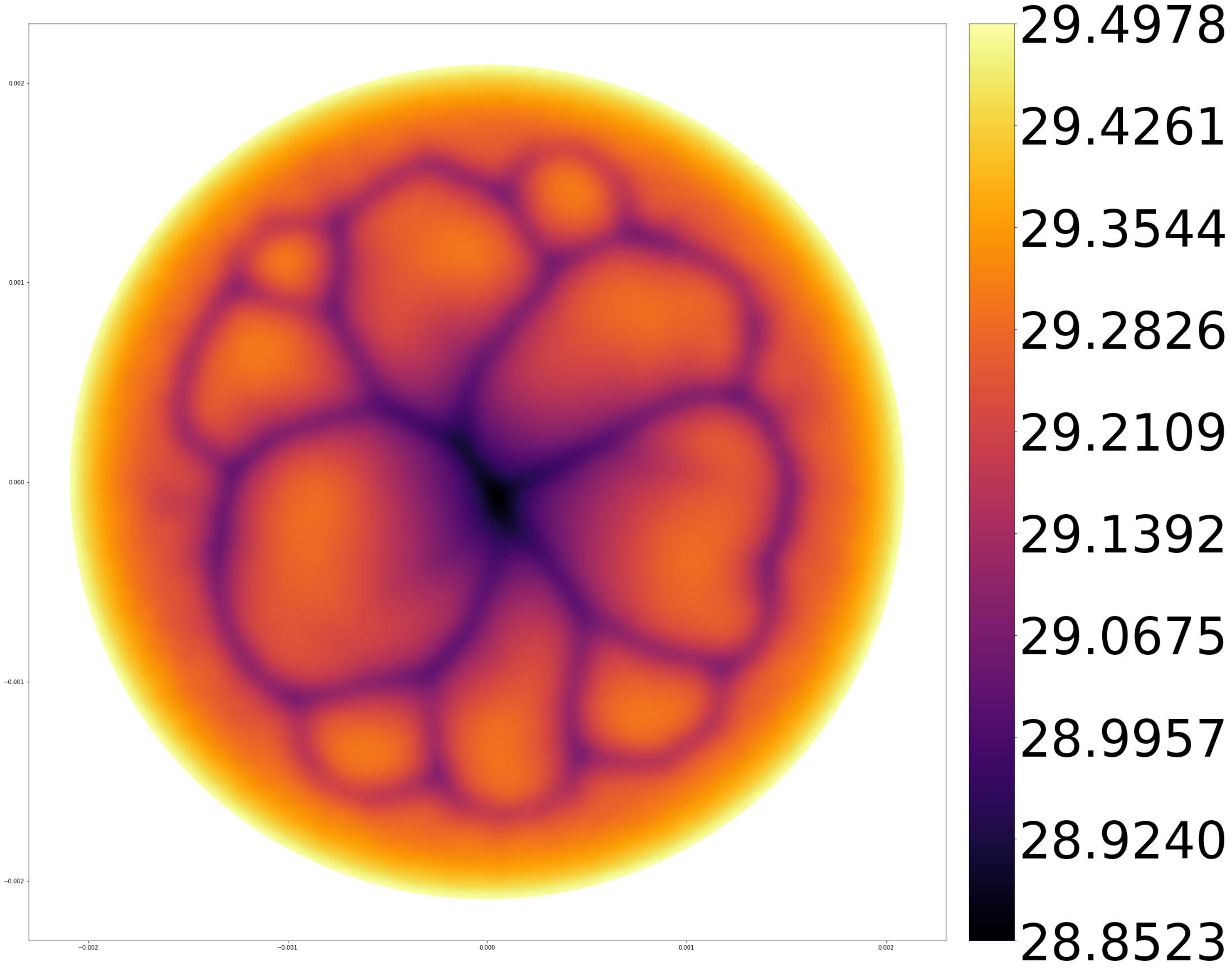}
\end{center}
\caption{Temporal dependence of the BM instabilities obtained for 
the silicone oil droplet for $\theta=20.49^\circ$.
The graphs correspond to the following values of the
evaporation time: 8, 18, 28, 38, 48, 58, 68 sec., respectively.
Temporal evolution of the droplet profile is not taken into account,
so the contact angle is fixed during the simulation.
}
\label{wang4efig}
\end{figure}

\begin{figure}[ht]
\begin{center}
\includegraphics[width=0.90\textwidth]{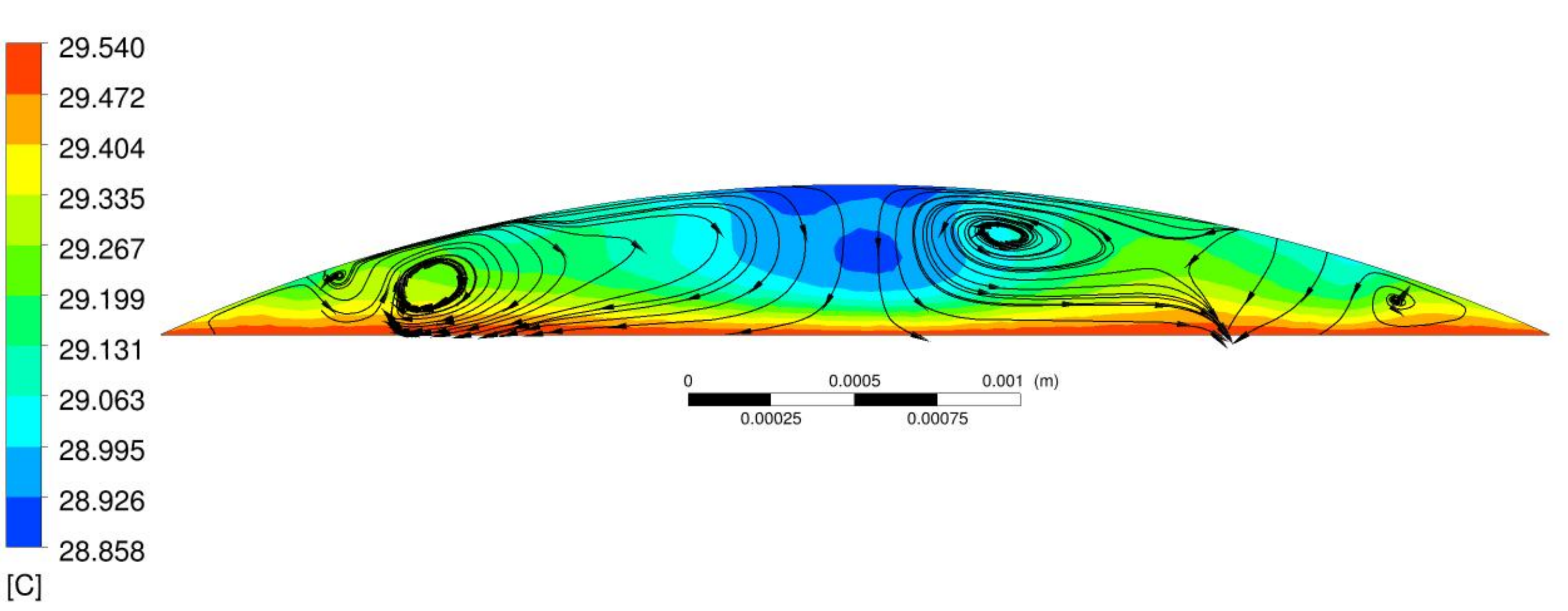}
\end{center}
\caption{Side view from the simulations of the silicone oil droplet with $\theta=20.49^\circ$.}
\label{wang4cside}
\end{figure}

\begin{figure}[ht]
\begin{center}
\includegraphics[width=0.36\textwidth]{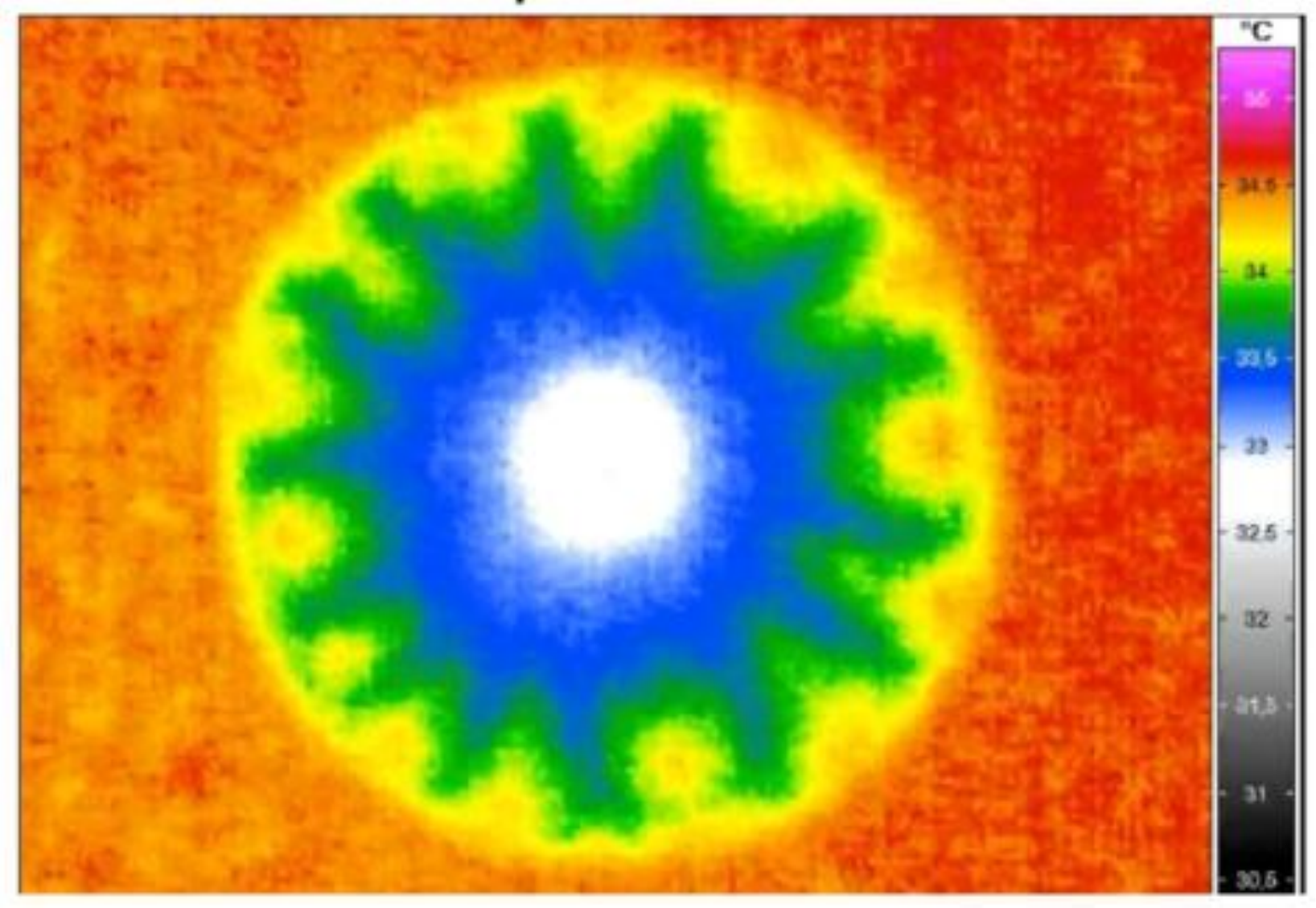}
\includegraphics[width=0.325\textwidth]{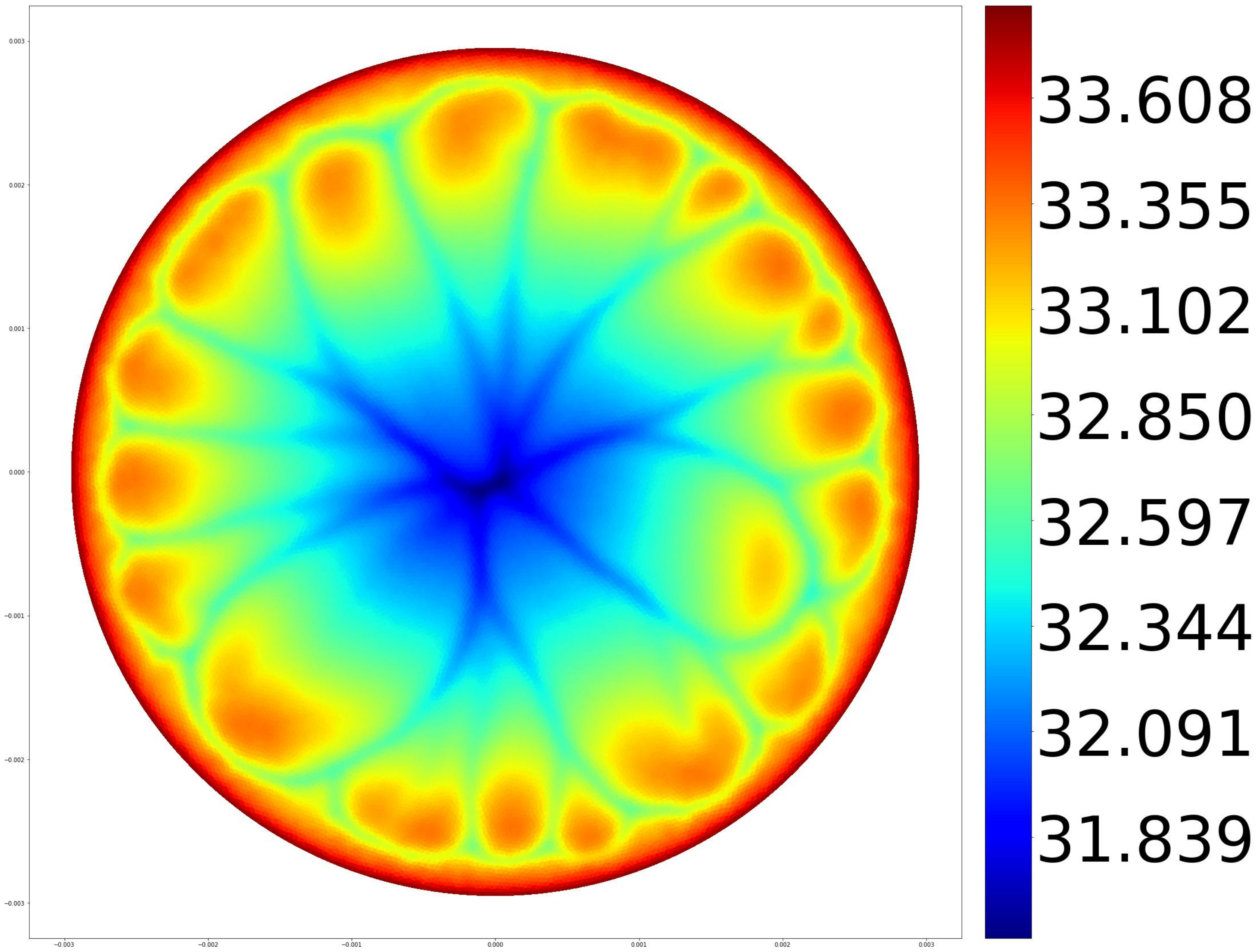}
\includegraphics[width=0.225\textwidth]{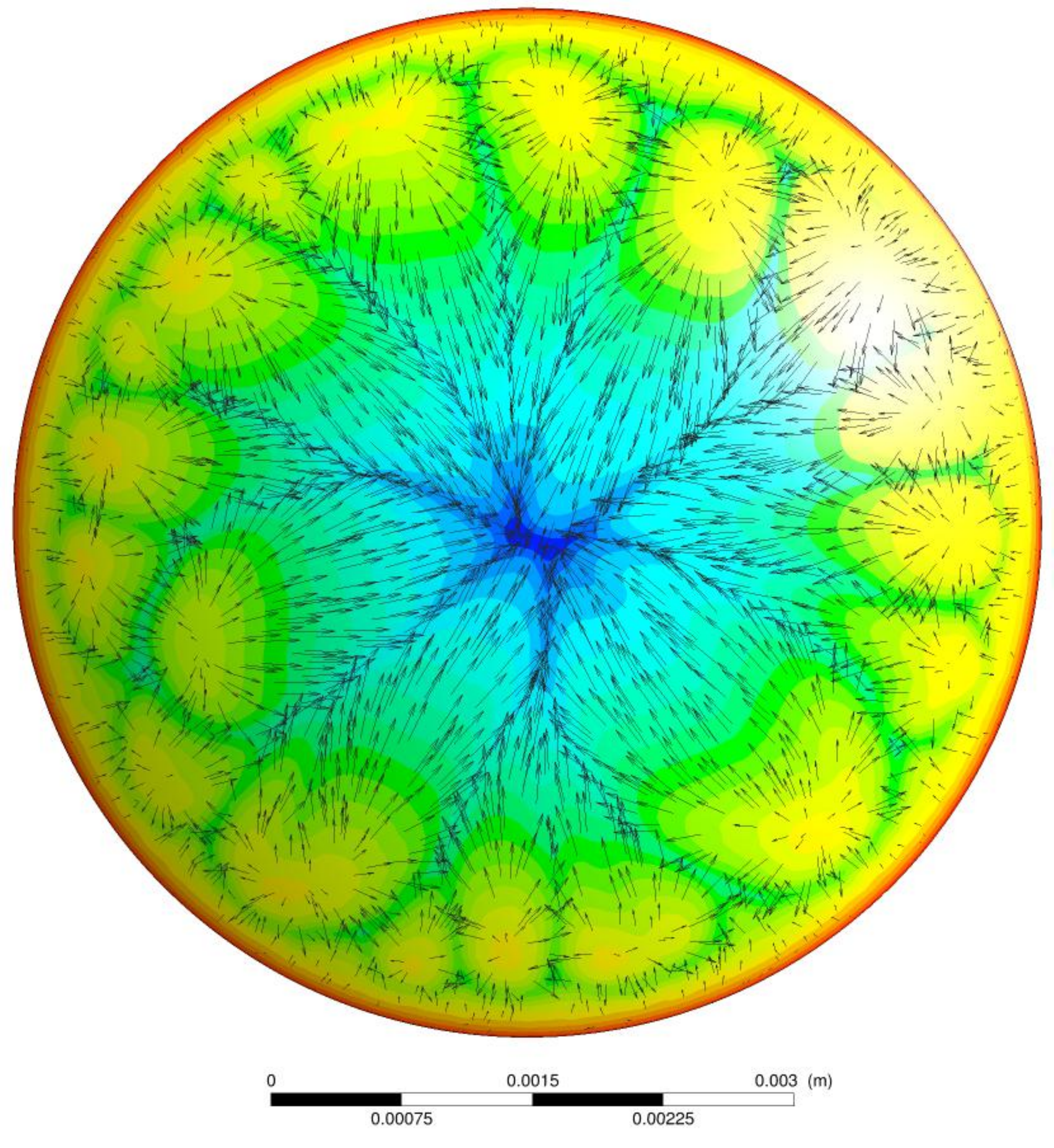}
\end{center}
\caption{ Surface temperature patterns and the velocity vector field plot
for the ethanol droplet with $\theta=29.2^\circ$. Left panel shows experimental data
and is reprinted with permission from Ref.~\cite{semenov2017}
(Copyright (2017) AIP Publishing).
Middle panel shows the corresponding surface temperature pattern
obtained by means of computer simulation
using the parameters from Table~\ref{NotationsTable}.
Right panel shows the vector field plot of numerically
obtained surface velocity.}
\label{sem2017fig}
\end{figure}

\begin{figure}[ht]
\begin{center}
\includegraphics[width=0.50\textwidth]{sem29num}
\includegraphics[width=0.48\textwidth]{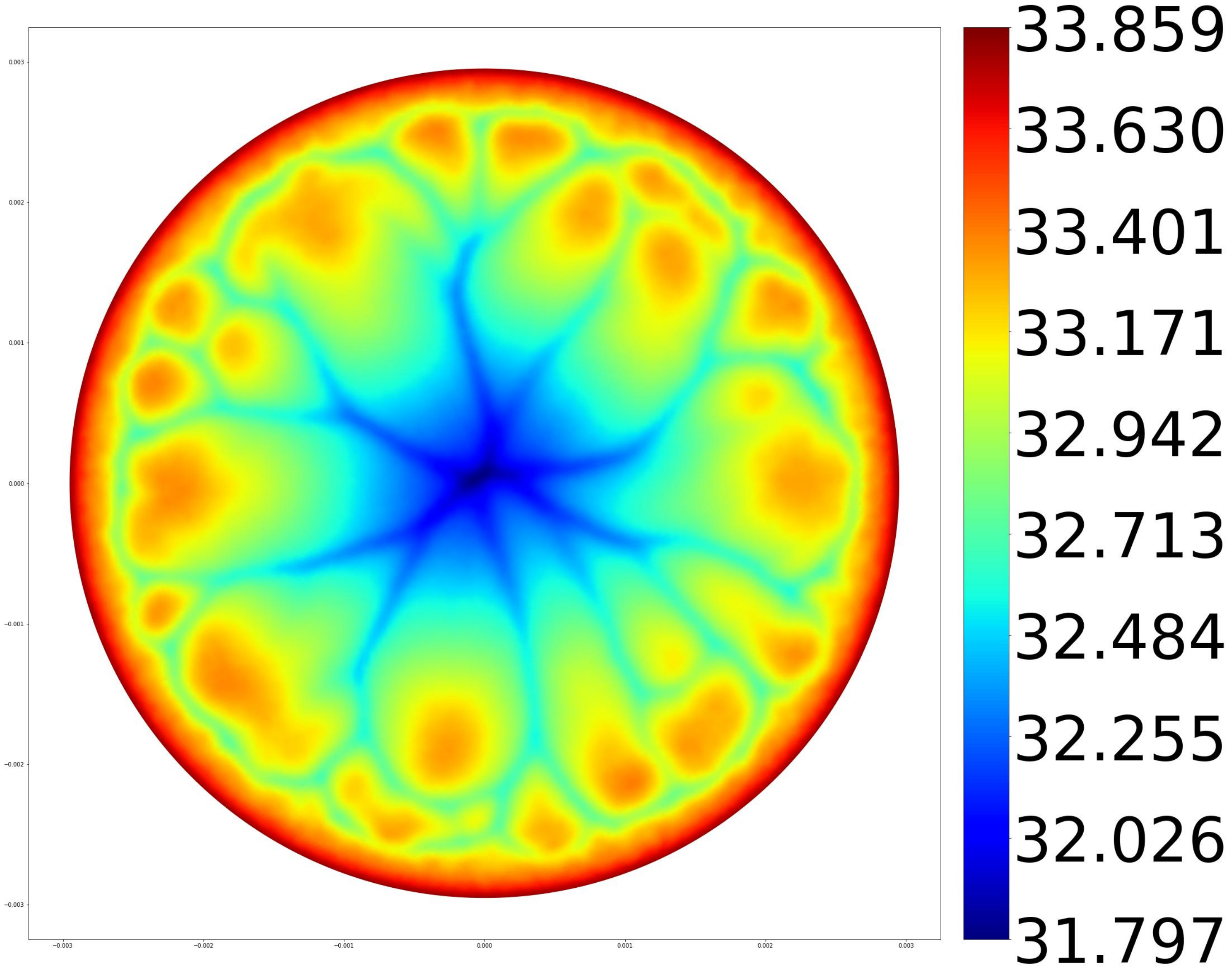}
\includegraphics[width=0.49\textwidth]{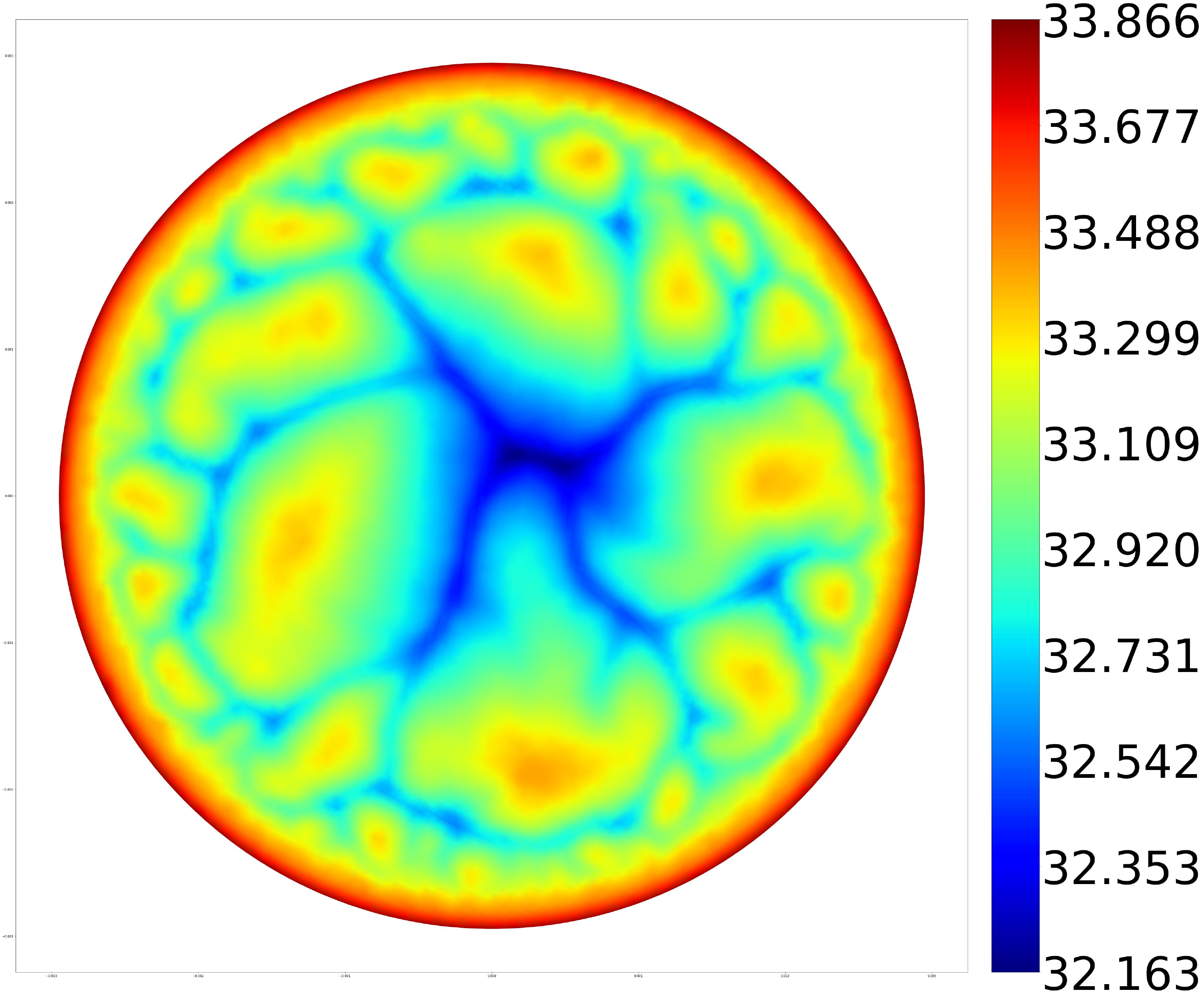}
\includegraphics[width=0.49\textwidth]{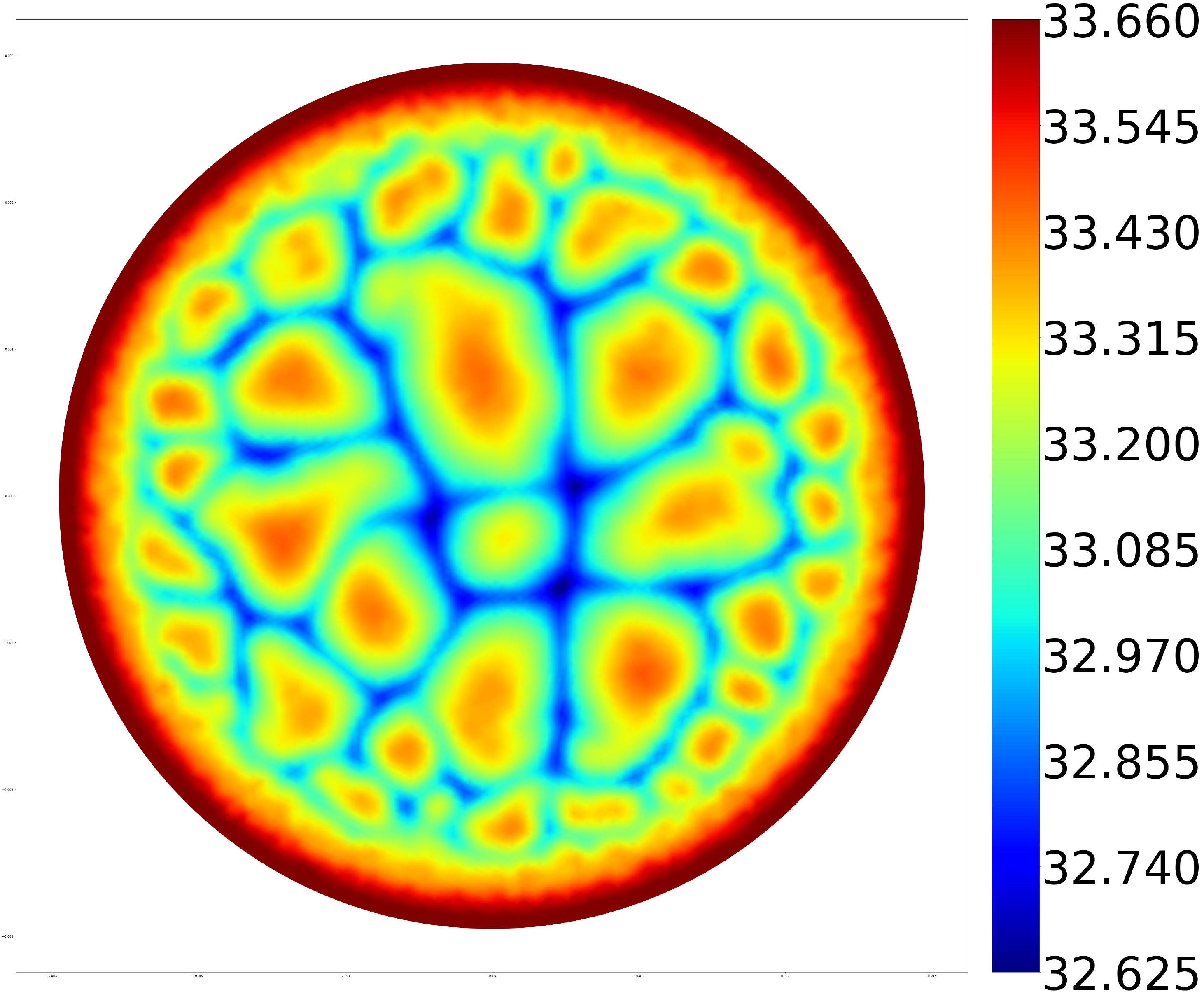}
\end{center}
\caption{Thermal patterns obtained by computer simulation of the ethanol droplet
for $\theta=29.2^\circ$, $\theta=24^\circ$, $\theta=20^\circ$ and $\theta=15^\circ$.}
\label{sem2017smaller}
\end{figure}

Fig.~\ref{wang2019fig} shows experimental results from Fig.4 in~\cite{wang2019}
and our numerical results for the drying droplet of $0.65$ cSt silicone oil.
Here $T_s=302.69$ K, $T_a=294.88$ K, $R=2.09$ mm, and the droplet contact
angle is $28.39^\circ$, $20.49^\circ$ and $17.18^\circ$.
The diffusion of vapor in air, the hydrodynamics in the drop, 
effects of the thermal conduction in all three phases and thermal radiation
have been taken into account as discussed above in Sec.~\ref{mathmodel}.

The obtained behavior of the BM cells qualitatively agrees with the experimental data:
there is a flower pattern for large contact angle, while the BM cells occupy the whole
droplet surface for small contact angles.
The simulation did not take into account the temporal dynamics
of the droplet profile, and so the quantitative details such as the particular
number of BM cells can vary.

We note that the Prandtl number is quite large for the problems
under consideration, so the HTW instability mechanism~\cite{smith1983,smith1986}
tends to move the BM cells from the droplet apex towards the contact line.

Fig.~\ref{wang4efig} shows that the oscillation of the pattern is very slow
compared to the characteristic frequences of HTWs in~\cite{sefiane2008,sefiane2010}.
Therefore, we confirm the existence of irregular oscillatory BM convection 
rather than HTWs in this case.
The Marangoni convection cells are sufficiently deep and are extended in the droplet bulk,
as seen in Fig.~\ref{wang4cside}.

Fig.~\ref{sem2017fig} was obtained for the drying droplet of ethanol.
Here, $T_s=307.05$ K, $T_a=297.55$ K, $R=2.95$ mm, $\theta=29.2^\circ$.
The temporal dynamics of the droplet profile has not been taken into account.
The numerical results qualitatively agree with the experimental data.
We obtained 20 BM cells, similarly to the numerical results of~\cite{semenov2017}
which were obtained by much different method.
We confirm the result of~\cite{semenov2017} that the obtained instability can be described
as unsteady BM convection rather than HTWs.

We did not calculate the IR image with Eq.(31) from~\cite{semenov2017} 
because we believe that equation is incorrect. In particular,
the equation is written under the assumption 
that the thermal radiation propagates only in the upright direction.
Also, the equation employs the Stefan-Bolzmann equation and does not take into account that
the maximum of Planck distribution is outside of the mid-wavelength range visible
by the IR camera, i.e., the IR camera passes by major part of the radiation.

Fig.~\ref{sem2017smaller} shows the obtained temperature distribution at the droplet surface
for $\theta=29.2^\circ$, $\theta=24^\circ$, $\theta=20^\circ$ and $\theta=15^\circ$.
There is a flower pattern for large contact angle, while the BM cells occupy the whole
droplet surface for sufficiently small contact angles.
This behavior is different from the one obtained numerically in~\cite{semenov2017}, where
the BM cells were located only in the vicinity of the contact line, and
the number of such BM cells increased with the decrease of the droplet height.
However, the behavior is similar to Fig.~\ref{wang2019fig}, which is natural
because the relevant physical parameters are quite close.

\section{Conclusion}
\label{ConclusionSec}

We have carried out three-dimensional computer simulations of unsteady 
Marangoni instabilities in drying sessile droplets of ethanol and silicone oil
of capillary size on a heated highly conductive substrate.
We have analyzed the relevance of physical effects in this case
and found that in addition to the nonstationary three-dimensional thermal conduction in the droplet, 
nonstationary three-dimensional dynamics of incompressible fluid and diffusion of vapor in air,
heat exchange in the vapor phase and radiation flux between
the drop and the environment also need to be taken into account.
However, contributions of such effects as nonstationary effects in the diffusion 
of vapor in air are insignificant.
Interestingly, we have found that the well-known analytical solution for the diffusion-limited 
evaporation rate~\cite{deegan2000}, that significantly simplifies the calculation of the heat 
transfer in the droplet, can also be employed for calculation of thermal conduction in the gas phase.

Our results agree well with the experimental results of~\cite{semenov2017,wang2019}.
We confirm the conclusion of these works that the instability in this case is a nonstationary 
BM convection rather than HTWs.

We also clarify the behavior of the BM instabilities with decrease of the contact angle.
We find that a decrease of the contact angle results in increasing BM cells, eventually
transforming the flower pattern into the spatial structure, where the BM cells 
occupy the whole droplet. Such a behavior has been observed experimentally in~\cite{wang2019}
but has not been obtained numerically until now, to the best of our knowledge.
We find that this behavior applies to the drying ethanol droplet of~\cite{semenov2017}
as well. This resolves the discrepancy between the BM cell behavior in~\cite{wang2019} and 
in~\cite{semenov2017}.

Our results are also consistent with the recent experimental results of~\cite{WangShi2020},
which show that the HTWs in the droplet appear only for relatively
large contact angles.

We believe that our results represent a useful step towards a better understanding 
of the thermocapillary instabilities in evaporating droplets.

\section*{Acknowledgments}

This work was supported by the Russian Science Foundation project No. 18-71-10061.

\bibliography{refs}

\begin{thebibliography}{39}%
\makeatletter
\providecommand \@ifxundefined [1]{%
 \@ifx{#1\undefined}
}%
\providecommand \@ifnum [1]{%
 \ifnum #1\expandafter \@firstoftwo
 \else \expandafter \@secondoftwo
 \fi
}%
\providecommand \@ifx [1]{%
 \ifx #1\expandafter \@firstoftwo
 \else \expandafter \@secondoftwo
 \fi
}%
\providecommand \natexlab [1]{#1}%
\providecommand \enquote  [1]{``#1''}%
\providecommand \bibnamefont  [1]{#1}%
\providecommand \bibfnamefont [1]{#1}%
\providecommand \citenamefont [1]{#1}%
\providecommand \href@noop [0]{\@secondoftwo}%
\providecommand \href [0]{\begingroup \@sanitize@url \@href}%
\providecommand \@href[1]{\@@startlink{#1}\@@href}%
\providecommand \@@href[1]{\endgroup#1\@@endlink}%
\providecommand \@sanitize@url [0]{\catcode `\\12\catcode `\$12\catcode
  `\&12\catcode `\#12\catcode `\^12\catcode `\_12\catcode `\%12\relax}%
\providecommand \@@startlink[1]{}%
\providecommand \@@endlink[0]{}%
\providecommand \url  [0]{\begingroup\@sanitize@url \@url }%
\providecommand \@url [1]{\endgroup\@href {#1}{\urlprefix }}%
\providecommand \urlprefix  [0]{URL }%
\providecommand \Eprint [0]{\href }%
\providecommand \doibase [0]{http://dx.doi.org/}%
\providecommand \selectlanguage [0]{\@gobble}%
\providecommand \bibinfo  [0]{\@secondoftwo}%
\providecommand \bibfield  [0]{\@secondoftwo}%
\providecommand \translation [1]{[#1]}%
\providecommand \BibitemOpen [0]{}%
\providecommand \bibitemStop [0]{}%
\providecommand \bibitemNoStop [0]{.\EOS\space}%
\providecommand \EOS [0]{\spacefactor3000\relax}%
\providecommand \BibitemShut  [1]{\csname bibitem#1\endcsname}%
\let\auto@bib@innerbib\@empty
\bibitem [{\citenamefont {Brutin}\ and\ \citenamefont
  {Starov}(2018)}]{brutin2018}%
  \BibitemOpen
  \bibfield  {author} {\bibinfo {author} {\bibfnamefont {D}~\bibnamefont
  {Brutin}}\ and\ \bibinfo {author} {\bibfnamefont {V}~\bibnamefont {Starov}},\
  }\bibfield  {title} {\enquote {\bibinfo {title} {Recent advances in droplet
  wetting and evaporation},}\ }\href {\doibase 10.1039/C6CS00902F} {\bibfield
  {journal} {\bibinfo  {journal} {Chem. Soc. Rev.}\ }\textbf {\bibinfo {volume}
  {47}},\ \bibinfo {pages} {558--585} (\bibinfo {year} {2018})}\BibitemShut
  {NoStop}%
\bibitem [{\citenamefont {Larson}(2014)}]{larson2014}%
  \BibitemOpen
  \bibfield  {author} {\bibinfo {author} {\bibfnamefont {Ronald~G}\
  \bibnamefont {Larson}},\ }\bibfield  {title} {\enquote {\bibinfo {title}
  {Transport and deposition patterns in drying sessile droplets},}\ }\href
  {\doibase 10.1002/aic.14338} {\bibfield  {journal} {\bibinfo  {journal}
  {AIChE J.}\ }\textbf {\bibinfo {volume} {60}},\ \bibinfo {pages} {1538--1571}
  (\bibinfo {year} {2014})}\BibitemShut {NoStop}%
\bibitem [{\citenamefont {Erbil}(2012)}]{erbil2012}%
  \BibitemOpen
  \bibfield  {author} {\bibinfo {author} {\bibfnamefont {H~Yildirim}\
  \bibnamefont {Erbil}},\ }\bibfield  {title} {\enquote {\bibinfo {title}
  {Evaporation of pure liquid sessile and spherical suspended drops: A
  review},}\ }\href {\doibase 10.1016/j.cis.2011.12.006} {\bibfield  {journal}
  {\bibinfo  {journal} {Adv. Colloid Interface Sci.}\ }\textbf {\bibinfo
  {volume} {170}},\ \bibinfo {pages} {67--86} (\bibinfo {year}
  {2012})}\BibitemShut {NoStop}%
\bibitem [{\citenamefont {Zang}\ \emph {et~al.}(2019)\citenamefont {Zang},
  \citenamefont {Tarafdar}, \citenamefont {Tarasevich}, \citenamefont {{Dutta
  Choudhury}},\ and\ \citenamefont {Dutta}}]{Zang2019}%
  \BibitemOpen
  \bibfield  {author} {\bibinfo {author} {\bibfnamefont {Duyang}\ \bibnamefont
  {Zang}}, \bibinfo {author} {\bibfnamefont {Sujata}\ \bibnamefont {Tarafdar}},
  \bibinfo {author} {\bibfnamefont {Yuri~Yu.}\ \bibnamefont {Tarasevich}},
  \bibinfo {author} {\bibfnamefont {Moutushi}\ \bibnamefont {{Dutta
  Choudhury}}}, \ and\ \bibinfo {author} {\bibfnamefont {Tapati}\ \bibnamefont
  {Dutta}},\ }\bibfield  {title} {\enquote {\bibinfo {title} {Evaporation of a
  droplet: From physics to applications},}\ }\href {\doibase
  10.1016/j.physrep.2019.01.008} {\bibfield  {journal} {\bibinfo  {journal}
  {Physics Reports}\ }\textbf {\bibinfo {volume} {804}},\ \bibinfo {pages}
  {1--56} (\bibinfo {year} {2019})}\BibitemShut {NoStop}%
\bibitem [{\citenamefont {Giorgiutti-Dauphin{\'e}}\ and\ \citenamefont
  {Pauchard}(2018)}]{Pauchard2018}%
  \BibitemOpen
  \bibfield  {author} {\bibinfo {author} {\bibfnamefont {F.}~\bibnamefont
  {Giorgiutti-Dauphin{\'e}}}\ and\ \bibinfo {author} {\bibfnamefont
  {L.}~\bibnamefont {Pauchard}},\ }\bibfield  {title} {\enquote {\bibinfo
  {title} {Drying drops},}\ }\href {\doibase 10.1140/epje/i2018-11639-2}
  {\bibfield  {journal} {\bibinfo  {journal} {Eur. Phys. J. E}\ }\textbf
  {\bibinfo {volume} {41}},\ \bibinfo {pages} {32} (\bibinfo {year}
  {2018})}\BibitemShut {NoStop}%
\bibitem [{\citenamefont {Shao}\ \emph {et~al.}(2020)\citenamefont {Shao},
  \citenamefont {Duan}, \citenamefont {Hou},\ and\ \citenamefont
  {Zhong}}]{Shao2020}%
  \BibitemOpen
  \bibfield  {author} {\bibinfo {author} {\bibfnamefont {Xiaoxiao}\
  \bibnamefont {Shao}}, \bibinfo {author} {\bibfnamefont {Fei}\ \bibnamefont
  {Duan}}, \bibinfo {author} {\bibfnamefont {Yu}~\bibnamefont {Hou}}, \ and\
  \bibinfo {author} {\bibfnamefont {Xin}\ \bibnamefont {Zhong}},\ }\bibfield
  {title} {\enquote {\bibinfo {title} {Role of surfactant in controlling the
  deposition pattern of a particle-laden droplet: Fundamentals and
  strategies},}\ }\href {\doibase 10.1016/j.cis.2019.102049} {\bibfield
  {journal} {\bibinfo  {journal} {Adv. Colloid Interface Sci.}\ }\textbf
  {\bibinfo {volume} {275}},\ \bibinfo {pages} {102049} (\bibinfo {year}
  {2020})}\BibitemShut {NoStop}%
\bibitem [{\citenamefont {Savino}\ and\ \citenamefont
  {Fico}(2004)}]{Savino2004}%
  \BibitemOpen
  \bibfield  {author} {\bibinfo {author} {\bibfnamefont {R.}~\bibnamefont
  {Savino}}\ and\ \bibinfo {author} {\bibfnamefont {S.}~\bibnamefont {Fico}},\
  }\bibfield  {title} {\enquote {\bibinfo {title} {Transient marangoni
  convection in hanging evaporating drops},}\ }\href {\doibase
  10.1063/1.1772380} {\bibfield  {journal} {\bibinfo  {journal} {Phys. Fluids}\
  }\textbf {\bibinfo {volume} {16}},\ \bibinfo {pages} {3738--3754} (\bibinfo
  {year} {2004})}\BibitemShut {NoStop}%
\bibitem [{\citenamefont {Kang}\ \emph {et~al.}(2004)\citenamefont {Kang},
  \citenamefont {Lee}, \citenamefont {Lee},\ and\ \citenamefont
  {Kang}}]{Kang2004}%
  \BibitemOpen
  \bibfield  {author} {\bibinfo {author} {\bibfnamefont {Kwan~Hyoung}\
  \bibnamefont {Kang}}, \bibinfo {author} {\bibfnamefont {Sang~Joon}\
  \bibnamefont {Lee}}, \bibinfo {author} {\bibfnamefont {Choung~Mook}\
  \bibnamefont {Lee}}, \ and\ \bibinfo {author} {\bibfnamefont {In~Seok}\
  \bibnamefont {Kang}},\ }\bibfield  {title} {\enquote {\bibinfo {title}
  {Quantitative visualization of flow inside an evaporating droplet using the
  ray tracing method},}\ }\href {\doibase 10.1088/0957-0233/15/6/009}
  {\bibfield  {journal} {\bibinfo  {journal} {Meas. Sci. Technol.}\ }\textbf
  {\bibinfo {volume} {15}},\ \bibinfo {pages} {1104--1112} (\bibinfo {year}
  {2004})}\BibitemShut {NoStop}%
\bibitem [{\citenamefont {Hu}\ and\ \citenamefont
  {Larson}(2006)}]{HuLarsonReversal}%
  \BibitemOpen
  \bibfield  {author} {\bibinfo {author} {\bibfnamefont {Hua}\ \bibnamefont
  {Hu}}\ and\ \bibinfo {author} {\bibfnamefont {Ronald~G.}\ \bibnamefont
  {Larson}},\ }\bibfield  {title} {\enquote {\bibinfo {title} {Marangoni effect
  reverses coffee-ring depositions},}\ }\href {\doibase 10.1021/jp0609232}
  {\bibfield  {journal} {\bibinfo  {journal} {J. Phys. Chem. B}\ }\textbf
  {\bibinfo {volume} {110}},\ \bibinfo {pages} {7090--7094} (\bibinfo {year}
  {2006})}\BibitemShut {NoStop}%
\bibitem [{\citenamefont {Ristenpart}\ \emph {et~al.}(2007)\citenamefont
  {Ristenpart}, \citenamefont {Kim}, \citenamefont {Domingues}, \citenamefont
  {Wan},\ and\ \citenamefont {Stone}}]{ristenpart2007}%
  \BibitemOpen
  \bibfield  {author} {\bibinfo {author} {\bibfnamefont {W.~D.}\ \bibnamefont
  {Ristenpart}}, \bibinfo {author} {\bibfnamefont {P.~G.}\ \bibnamefont {Kim}},
  \bibinfo {author} {\bibfnamefont {C.}~\bibnamefont {Domingues}}, \bibinfo
  {author} {\bibfnamefont {J.}~\bibnamefont {Wan}}, \ and\ \bibinfo {author}
  {\bibfnamefont {H.~A.}\ \bibnamefont {Stone}},\ }\bibfield  {title} {\enquote
  {\bibinfo {title} {Influence of substrate conductivity on circulation
  reversal in evaporating drops},}\ }\href {\doibase
  10.1103/PhysRevLett.99.234502} {\bibfield  {journal} {\bibinfo  {journal}
  {Phys. Rev. Lett.}\ }\textbf {\bibinfo {volume} {99}},\ \bibinfo {pages}
  {234502} (\bibinfo {year} {2007})}\BibitemShut {NoStop}%
\bibitem [{\citenamefont {Sefiane}\ \emph {et~al.}(2008)\citenamefont
  {Sefiane}, \citenamefont {Moffat}, \citenamefont {Matar},\ and\ \citenamefont
  {Craster}}]{sefiane2008}%
  \BibitemOpen
  \bibfield  {author} {\bibinfo {author} {\bibfnamefont {K.}~\bibnamefont
  {Sefiane}}, \bibinfo {author} {\bibfnamefont {J.~R.}\ \bibnamefont {Moffat}},
  \bibinfo {author} {\bibfnamefont {O.~K.}\ \bibnamefont {Matar}}, \ and\
  \bibinfo {author} {\bibfnamefont {R.~V.}\ \bibnamefont {Craster}},\
  }\bibfield  {title} {\enquote {\bibinfo {title} {Self-excited hydrothermal
  waves in evaporating sessile drops},}\ }\href {\doibase 10.1063/1.2969072}
  {\bibfield  {journal} {\bibinfo  {journal} {Appl. Phys. Lett.}\ }\textbf
  {\bibinfo {volume} {93}},\ \bibinfo {pages} {074103} (\bibinfo {year}
  {2008})}\BibitemShut {NoStop}%
\bibitem [{\citenamefont {Sefiane}\ \emph {et~al.}(2010)\citenamefont
  {Sefiane}, \citenamefont {Steinchen},\ and\ \citenamefont
  {Moffat}}]{sefiane2010}%
  \BibitemOpen
  \bibfield  {author} {\bibinfo {author} {\bibfnamefont {K.}~\bibnamefont
  {Sefiane}}, \bibinfo {author} {\bibfnamefont {A.}~\bibnamefont {Steinchen}},
  \ and\ \bibinfo {author} {\bibfnamefont {R.}~\bibnamefont {Moffat}},\
  }\bibfield  {title} {\enquote {\bibinfo {title} {On hydrothermal waves
  observed during evaporation of sessile droplets},}\ }\href {\doibase
  10.1016/j.colsurfa.2010.02.015} {\bibfield  {journal} {\bibinfo  {journal}
  {Colloids Surf. A}\ }\textbf {\bibinfo {volume} {365}},\ \bibinfo {pages}
  {95--108} (\bibinfo {year} {2010})}\BibitemShut {NoStop}%
\bibitem [{\citenamefont {Brutin}\ \emph {et~al.}(2011)\citenamefont {Brutin},
  \citenamefont {Sobac}, \citenamefont {Rigollet},\ and\ \citenamefont
  {Niliot]}}]{brutin2011}%
  \BibitemOpen
  \bibfield  {author} {\bibinfo {author} {\bibfnamefont {D.}~\bibnamefont
  {Brutin}}, \bibinfo {author} {\bibfnamefont {B.}~\bibnamefont {Sobac}},
  \bibinfo {author} {\bibfnamefont {F.}~\bibnamefont {Rigollet}}, \ and\
  \bibinfo {author} {\bibfnamefont {C.~[Le}\ \bibnamefont {Niliot]}},\
  }\bibfield  {title} {\enquote {\bibinfo {title} {Infrared visualization of
  thermal motion inside a sessile drop deposited onto a heated surface},}\
  }\href {\doibase 10.1016/j.expthermflusci.2010.12.004} {\bibfield  {journal}
  {\bibinfo  {journal} {Exp. Therm Fluid Sci.}\ }\textbf {\bibinfo {volume}
  {35}},\ \bibinfo {pages} {521--530} (\bibinfo {year} {2011})}\BibitemShut
  {NoStop}%
\bibitem [{\citenamefont {Sobac}\ and\ \citenamefont
  {Brutin}(2012)}]{sobac2012}%
  \BibitemOpen
  \bibfield  {author} {\bibinfo {author} {\bibfnamefont {B.}~\bibnamefont
  {Sobac}}\ and\ \bibinfo {author} {\bibfnamefont {D.}~\bibnamefont {Brutin}},\
  }\bibfield  {title} {\enquote {\bibinfo {title} {Thermocapillary
  instabilities in an evaporating drop deposited onto a heated substrate},}\
  }\href {\doibase 10.1063/1.3692267} {\bibfield  {journal} {\bibinfo
  {journal} {Phys. Fluids}\ }\textbf {\bibinfo {volume} {24}},\ \bibinfo
  {pages} {032103} (\bibinfo {year} {2012})}\BibitemShut {NoStop}%
\bibitem [{\citenamefont {Carle}\ \emph {et~al.}(2012)\citenamefont {Carle},
  \citenamefont {Sobac},\ and\ \citenamefont {Brutin}}]{carle2012}%
  \BibitemOpen
  \bibfield  {author} {\bibinfo {author} {\bibfnamefont {F.}~\bibnamefont
  {Carle}}, \bibinfo {author} {\bibfnamefont {B.}~\bibnamefont {Sobac}}, \ and\
  \bibinfo {author} {\bibfnamefont {D.}~\bibnamefont {Brutin}},\ }\bibfield
  {title} {\enquote {\bibinfo {title} {Hydrothermal waves on ethanol droplets
  evaporating under terrestrial and reduced gravity levels},}\ }\href {\doibase
  10.1017/jfm.2012.446} {\bibfield  {journal} {\bibinfo  {journal} {J. Fluid
  Mech.}\ }\textbf {\bibinfo {volume} {712}},\ \bibinfo {pages} {614--623}
  (\bibinfo {year} {2012})}\BibitemShut {NoStop}%
\bibitem [{\citenamefont {Sefiane}\ \emph {et~al.}(2013)\citenamefont
  {Sefiane}, \citenamefont {Fukatani}, \citenamefont {Takata},\ and\
  \citenamefont {Kim}}]{sefiane2013}%
  \BibitemOpen
  \bibfield  {author} {\bibinfo {author} {\bibfnamefont {Khellil}\ \bibnamefont
  {Sefiane}}, \bibinfo {author} {\bibfnamefont {Yuki}\ \bibnamefont
  {Fukatani}}, \bibinfo {author} {\bibfnamefont {Yasuyuki}\ \bibnamefont
  {Takata}}, \ and\ \bibinfo {author} {\bibfnamefont {Jungho}\ \bibnamefont
  {Kim}},\ }\bibfield  {title} {\enquote {\bibinfo {title} {Thermal patterns
  and hydrothermal waves (htws) in volatile drops},}\ }\href {\doibase
  10.1021/la402247n} {\bibfield  {journal} {\bibinfo  {journal} {Langmuir}\
  }\textbf {\bibinfo {volume} {29}},\ \bibinfo {pages} {9750--9760} (\bibinfo
  {year} {2013})}\BibitemShut {NoStop}%
\bibitem [{\citenamefont {Karapetsas}\ \emph {et~al.}(2012)\citenamefont
  {Karapetsas}, \citenamefont {Matar}, \citenamefont {Valluri},\ and\
  \citenamefont {Sefiane}}]{karapetsas2012}%
  \BibitemOpen
  \bibfield  {author} {\bibinfo {author} {\bibfnamefont {George}\ \bibnamefont
  {Karapetsas}}, \bibinfo {author} {\bibfnamefont {Omar~K.}\ \bibnamefont
  {Matar}}, \bibinfo {author} {\bibfnamefont {Prashant}\ \bibnamefont
  {Valluri}}, \ and\ \bibinfo {author} {\bibfnamefont {Khellil}\ \bibnamefont
  {Sefiane}},\ }\bibfield  {title} {\enquote {\bibinfo {title} {Convective
  rolls and hydrothermal waves in evaporating sessile drops},}\ }\href
  {\doibase 10.1021/la3019088} {\bibfield  {journal} {\bibinfo  {journal}
  {Langmuir}\ }\textbf {\bibinfo {volume} {28}},\ \bibinfo {pages}
  {11433--11439} (\bibinfo {year} {2012})}\BibitemShut {NoStop}%
\bibitem [{\citenamefont {S\'aenz}\ \emph {et~al.}(2014)\citenamefont
  {S\'aenz}, \citenamefont {Valluri}, \citenamefont {Sefiane}, \citenamefont
  {Karapetsas},\ and\ \citenamefont {Matar}}]{saenz2014}%
  \BibitemOpen
  \bibfield  {author} {\bibinfo {author} {\bibfnamefont {P.~J.}\ \bibnamefont
  {S\'aenz}}, \bibinfo {author} {\bibfnamefont {P.}~\bibnamefont {Valluri}},
  \bibinfo {author} {\bibfnamefont {K.}~\bibnamefont {Sefiane}}, \bibinfo
  {author} {\bibfnamefont {G.}~\bibnamefont {Karapetsas}}, \ and\ \bibinfo
  {author} {\bibfnamefont {O.~K.}\ \bibnamefont {Matar}},\ }\bibfield  {title}
  {\enquote {\bibinfo {title} {On phase change in marangoni-driven flows and
  its effects on the hydrothermal-wave instabilities},}\ }\href {\doibase
  10.1063/1.4866770} {\bibfield  {journal} {\bibinfo  {journal} {Phys. Fluids}\
  }\textbf {\bibinfo {volume} {26}},\ \bibinfo {pages} {024114} (\bibinfo
  {year} {2014})}\BibitemShut {NoStop}%
\bibitem [{\citenamefont {Zhu}\ \emph {et~al.}(2019)\citenamefont {Zhu},
  \citenamefont {Shi},\ and\ \citenamefont {Feng}}]{zhu2019}%
  \BibitemOpen
  \bibfield  {author} {\bibinfo {author} {\bibfnamefont {Ji-Long}\ \bibnamefont
  {Zhu}}, \bibinfo {author} {\bibfnamefont {Wan-Yuan}\ \bibnamefont {Shi}}, \
  and\ \bibinfo {author} {\bibfnamefont {Lin}\ \bibnamefont {Feng}},\
  }\bibfield  {title} {\enquote {\bibinfo {title} {B\'enard-marangoni
  instability in sessile droplet evaporating at constant contact angle mode on
  heated substrate},}\ }\href {\doibase
  https://doi.org/10.1016/j.ijheatmasstransfer.2019.01.082} {\bibfield
  {journal} {\bibinfo  {journal} {Int. J. Heat Mass Transfer}\ }\textbf
  {\bibinfo {volume} {134}},\ \bibinfo {pages} {784--795} (\bibinfo {year}
  {2019})}\BibitemShut {NoStop}%
\bibitem [{\citenamefont {Smith}\ and\ \citenamefont
  {Davis}(1983)}]{smith1983}%
  \BibitemOpen
  \bibfield  {author} {\bibinfo {author} {\bibfnamefont {Marc~K.}\ \bibnamefont
  {Smith}}\ and\ \bibinfo {author} {\bibfnamefont {Stephen~H.}\ \bibnamefont
  {Davis}},\ }\bibfield  {title} {\enquote {\bibinfo {title} {Instabilities of
  dynamic thermocapillary liquid layers. part 1. convective instabilities},}\
  }\href {\doibase 10.1017/S0022112083001512} {\bibfield  {journal} {\bibinfo
  {journal} {J. Fluid Mech.}\ }\textbf {\bibinfo {volume} {132}},\ \bibinfo
  {pages} {119--144} (\bibinfo {year} {1983})}\BibitemShut {NoStop}%
\bibitem [{\citenamefont {Smith}(1986)}]{smith1986}%
  \BibitemOpen
  \bibfield  {author} {\bibinfo {author} {\bibfnamefont {Marc~K.}\ \bibnamefont
  {Smith}},\ }\bibfield  {title} {\enquote {\bibinfo {title} {Instability
  mechanisms in dynamic thermocapillary liquid layers},}\ }\href {\doibase
  10.1063/1.865836} {\bibfield  {journal} {\bibinfo  {journal} {Phys. Fluids}\
  }\textbf {\bibinfo {volume} {29}},\ \bibinfo {pages} {3182--3186} (\bibinfo
  {year} {1986})}\BibitemShut {NoStop}%
\bibitem [{\citenamefont {Schwabe}\ \emph {et~al.}(1992)\citenamefont
  {Schwabe}, \citenamefont {M\"oller}, \citenamefont {Schneider},\ and\
  \citenamefont {Scharmann}}]{schwabe1992}%
  \BibitemOpen
  \bibfield  {author} {\bibinfo {author} {\bibfnamefont {D.}~\bibnamefont
  {Schwabe}}, \bibinfo {author} {\bibfnamefont {U.}~\bibnamefont {M\"oller}},
  \bibinfo {author} {\bibfnamefont {J.}~\bibnamefont {Schneider}}, \ and\
  \bibinfo {author} {\bibfnamefont {A.}~\bibnamefont {Scharmann}},\ }\bibfield
  {title} {\enquote {\bibinfo {title} {Instabilities of shallow dynamic
  thermocapillary liquid layers},}\ }\href {\doibase 10.1063/1.858478}
  {\bibfield  {journal} {\bibinfo  {journal} {Phys. Fluids A}\ }\textbf
  {\bibinfo {volume} {4}},\ \bibinfo {pages} {2368--2381} (\bibinfo {year}
  {1992})}\BibitemShut {NoStop}%
\bibitem [{\citenamefont {Riley}\ and\ \citenamefont
  {Neitzel}(1998)}]{riley1998}%
  \BibitemOpen
  \bibfield  {author} {\bibinfo {author} {\bibfnamefont {R.~J.}\ \bibnamefont
  {Riley}}\ and\ \bibinfo {author} {\bibfnamefont {G.~P.}\ \bibnamefont
  {Neitzel}},\ }\bibfield  {title} {\enquote {\bibinfo {title} {Instability of
  thermocapillary-buoyancy convection in shallow layers. part 1.
  characterization of steady and oscillatory instabilities},}\ }\href {\doibase
  10.1017/S0022112097008343} {\bibfield  {journal} {\bibinfo  {journal} {J.
  Fluid Mech.}\ }\textbf {\bibinfo {volume} {359}},\ \bibinfo {pages}
  {143--164} (\bibinfo {year} {1998})}\BibitemShut {NoStop}%
\bibitem [{\citenamefont {Shi}\ \emph {et~al.}(2017)\citenamefont {Shi},
  \citenamefont {Tang}, \citenamefont {Ma}, \citenamefont {Jia}, \citenamefont
  {Li},\ and\ \citenamefont {Feng}}]{shi2017}%
  \BibitemOpen
  \bibfield  {author} {\bibinfo {author} {\bibfnamefont {Wan-Yuan}\
  \bibnamefont {Shi}}, \bibinfo {author} {\bibfnamefont {Kai-Yi}\ \bibnamefont
  {Tang}}, \bibinfo {author} {\bibfnamefont {Jia-Nan}\ \bibnamefont {Ma}},
  \bibinfo {author} {\bibfnamefont {Yi-Wei}\ \bibnamefont {Jia}}, \bibinfo
  {author} {\bibfnamefont {Han-Ming}\ \bibnamefont {Li}}, \ and\ \bibinfo
  {author} {\bibfnamefont {Lin}\ \bibnamefont {Feng}},\ }\bibfield  {title}
  {\enquote {\bibinfo {title} {Marangoni convection instability in a sessile
  droplet with low volatility on heated substrate},}\ }\href {\doibase
  10.1016/j.ijthermalsci.2017.04.007} {\bibfield  {journal} {\bibinfo
  {journal} {Int. J. Therm. Sci.}\ }\textbf {\bibinfo {volume} {117}},\
  \bibinfo {pages} {274--286} (\bibinfo {year} {2017})}\BibitemShut {NoStop}%
\bibitem [{\citenamefont {Semenov}\ \emph
  {et~al.}(2017{\natexlab{a}})\citenamefont {Semenov}, \citenamefont {Carle},
  \citenamefont {Medale},\ and\ \citenamefont {Brutin}}]{semenov2017}%
  \BibitemOpen
  \bibfield  {author} {\bibinfo {author} {\bibfnamefont {Sergey}\ \bibnamefont
  {Semenov}}, \bibinfo {author} {\bibfnamefont {Florian}\ \bibnamefont
  {Carle}}, \bibinfo {author} {\bibfnamefont {Marc}\ \bibnamefont {Medale}}, \
  and\ \bibinfo {author} {\bibfnamefont {David}\ \bibnamefont {Brutin}},\
  }\bibfield  {title} {\enquote {\bibinfo {title} {3d unsteady computations of
  evaporative instabilities in a sessile drop of ethanol on a heated
  substrate},}\ }\href {\doibase 10.1063/1.5006707} {\bibfield  {journal}
  {\bibinfo  {journal} {Appl. Phys. Lett.}\ }\textbf {\bibinfo {volume}
  {111}},\ \bibinfo {pages} {241602} (\bibinfo {year}
  {2017}{\natexlab{a}})}\BibitemShut {NoStop}%
\bibitem [{\citenamefont {Semenov}\ \emph
  {et~al.}(2017{\natexlab{b}})\citenamefont {Semenov}, \citenamefont {Carle},
  \citenamefont {Medale},\ and\ \citenamefont {Brutin}}]{semenov2017PRE}%
  \BibitemOpen
  \bibfield  {author} {\bibinfo {author} {\bibfnamefont {Sergey}\ \bibnamefont
  {Semenov}}, \bibinfo {author} {\bibfnamefont {Florian}\ \bibnamefont
  {Carle}}, \bibinfo {author} {\bibfnamefont {Marc}\ \bibnamefont {Medale}}, \
  and\ \bibinfo {author} {\bibfnamefont {David}\ \bibnamefont {Brutin}},\
  }\bibfield  {title} {\enquote {\bibinfo {title} {Boundary conditions for a
  one-sided numerical model of evaporative instabilities in sessile drops of
  ethanol on heated substrates},}\ }\href {\doibase 10.1103/PhysRevE.96.063113}
  {\bibfield  {journal} {\bibinfo  {journal} {Phys. Rev. E}\ }\textbf {\bibinfo
  {volume} {96}},\ \bibinfo {pages} {063113} (\bibinfo {year}
  {2017}{\natexlab{b}})}\BibitemShut {NoStop}%
\bibitem [{\citenamefont {Wang}\ and\ \citenamefont {Shi}(2019)}]{wang2019}%
  \BibitemOpen
  \bibfield  {author} {\bibinfo {author} {\bibfnamefont {Tian-Shi}\
  \bibnamefont {Wang}}\ and\ \bibinfo {author} {\bibfnamefont {Wan-Yuan}\
  \bibnamefont {Shi}},\ }\bibfield  {title} {\enquote {\bibinfo {title}
  {Influence of substrate temperature on marangoni convection instabilities in
  a sessile droplet evaporating at constant contact line mode},}\ }\href
  {\doibase 10.1016/j.ijheatmasstransfer.2018.11.155} {\bibfield  {journal}
  {\bibinfo  {journal} {Int. J. Heat Mass Transfer}\ }\textbf {\bibinfo
  {volume} {131}},\ \bibinfo {pages} {1270--1278} (\bibinfo {year}
  {2019})}\BibitemShut {NoStop}%
\bibitem [{\citenamefont {Barash}\ \emph {et~al.}(2009)\citenamefont {Barash},
  \citenamefont {Bigioni}, \citenamefont {Vinokur},\ and\ \citenamefont
  {Shchur}}]{barash2009}%
  \BibitemOpen
  \bibfield  {author} {\bibinfo {author} {\bibfnamefont {L.~Yu.}\ \bibnamefont
  {Barash}}, \bibinfo {author} {\bibfnamefont {T.~P.}\ \bibnamefont {Bigioni}},
  \bibinfo {author} {\bibfnamefont {V.~M.}\ \bibnamefont {Vinokur}}, \ and\
  \bibinfo {author} {\bibfnamefont {L.~N.}\ \bibnamefont {Shchur}},\ }\bibfield
   {title} {\enquote {\bibinfo {title} {Evaporation and fluid dynamics of a
  sessile drop of capillary size},}\ }\href {\doibase
  10.1103/PhysRevE.79.046301} {\bibfield  {journal} {\bibinfo  {journal} {Phys.
  Rev. E}\ }\textbf {\bibinfo {volume} {79}},\ \bibinfo {pages} {046301}
  (\bibinfo {year} {2009})}\BibitemShut {NoStop}%
\bibitem [{\citenamefont {Lebedev}(1972)}]{lebedev}%
  \BibitemOpen
  \bibfield  {author} {\bibinfo {author} {\bibfnamefont {N.N.}\ \bibnamefont
  {Lebedev}},\ }\href@noop {} {\emph {\bibinfo {title} {Special Functions and
  Their Applications}}},\ Dover Books on Mathematics\ (\bibinfo  {publisher}
  {Dover Publications},\ \bibinfo {year} {1972})\BibitemShut {NoStop}%
\bibitem [{\citenamefont {Deegan}\ \emph {et~al.}(2000)\citenamefont {Deegan},
  \citenamefont {Bakajin}, \citenamefont {Dupont}, \citenamefont {Huber},
  \citenamefont {Nagel},\ and\ \citenamefont {Witten}}]{deegan2000}%
  \BibitemOpen
  \bibfield  {author} {\bibinfo {author} {\bibfnamefont {Robert~D.}\
  \bibnamefont {Deegan}}, \bibinfo {author} {\bibfnamefont {Olgica}\
  \bibnamefont {Bakajin}}, \bibinfo {author} {\bibfnamefont {Todd~F.}\
  \bibnamefont {Dupont}}, \bibinfo {author} {\bibfnamefont {Greg}\ \bibnamefont
  {Huber}}, \bibinfo {author} {\bibfnamefont {Sidney~R.}\ \bibnamefont
  {Nagel}}, \ and\ \bibinfo {author} {\bibfnamefont {Thomas~A.}\ \bibnamefont
  {Witten}},\ }\bibfield  {title} {\enquote {\bibinfo {title} {Contact line
  deposits in an evaporating drop},}\ }\href {\doibase 10.1103/PhysRevE.62.756}
  {\bibfield  {journal} {\bibinfo  {journal} {Phys. Rev. E}\ }\textbf {\bibinfo
  {volume} {62}},\ \bibinfo {pages} {756--765} (\bibinfo {year}
  {2000})}\BibitemShut {NoStop}%
\bibitem [{\citenamefont {Hu}\ and\ \citenamefont {Larson}(2002)}]{hu2002}%
  \BibitemOpen
  \bibfield  {author} {\bibinfo {author} {\bibfnamefont {Hua}\ \bibnamefont
  {Hu}}\ and\ \bibinfo {author} {\bibfnamefont {Ronald~G.}\ \bibnamefont
  {Larson}},\ }\bibfield  {title} {\enquote {\bibinfo {title} {Evaporation of a
  sessile droplet on a substrate},}\ }\href {\doibase 10.1021/jp0118322}
  {\bibfield  {journal} {\bibinfo  {journal} {J. Phys. Chem. B}\ }\textbf
  {\bibinfo {volume} {106}},\ \bibinfo {pages} {1334--1344} (\bibinfo {year}
  {2002})}\BibitemShut {NoStop}%
\bibitem [{\citenamefont {Ambrose}\ and\ \citenamefont
  {Sprake}(1970)}]{Ambrose1970}%
  \BibitemOpen
  \bibfield  {author} {\bibinfo {author} {\bibfnamefont {D.}~\bibnamefont
  {Ambrose}}\ and\ \bibinfo {author} {\bibfnamefont {C.H.S.}\ \bibnamefont
  {Sprake}},\ }\bibfield  {title} {\enquote {\bibinfo {title} {Thermodynamic
  properties of organic oxygen compounds xxv. vapour pressures and normal
  boiling temperatures of aliphatic alcohols},}\ }\href {\doibase
  https://doi.org/10.1016/0021-9614(70)90038-8} {\bibfield  {journal} {\bibinfo
   {journal} {J. Chem. Thermodyn.}\ }\textbf {\bibinfo {volume} {2}},\ \bibinfo
  {pages} {631--645} (\bibinfo {year} {1970})}\BibitemShut {NoStop}%
\bibitem [{\citenamefont {Fuchs}(1959)}]{fuchs1959}%
  \BibitemOpen
  \bibfield  {author} {\bibinfo {author} {\bibfnamefont {Nikolai~Albertovich}\
  \bibnamefont {Fuchs}},\ }\href@noop {} {\emph {\bibinfo {title} {Evaporation
  and droplet growth in gaseous media}}}\ (\bibinfo  {publisher} {Pergamon
  Press, Oxford},\ \bibinfo {year} {1959})\BibitemShut {NoStop}%
\bibitem [{\citenamefont {Landau}\ and\ \citenamefont {Lifshitz}(1982)}]{LL6}%
  \BibitemOpen
  \bibfield  {author} {\bibinfo {author} {\bibfnamefont {L.D.}\ \bibnamefont
  {Landau}}\ and\ \bibinfo {author} {\bibfnamefont {E.M.}\ \bibnamefont
  {Lifshitz}},\ }\href@noop {} {\emph {\bibinfo {title} {Course of Theoretical
  Physics VI: Fluid Mechanics}}}\ (\bibinfo  {publisher} {Pergamon Press,
  Oxford},\ \bibinfo {year} {1982})\BibitemShut {NoStop}%
\bibitem [{\citenamefont {Kattawar}\ and\ \citenamefont
  {Eisner}(1970)}]{kattawar1970}%
  \BibitemOpen
  \bibfield  {author} {\bibinfo {author} {\bibfnamefont {G.~W.}\ \bibnamefont
  {Kattawar}}\ and\ \bibinfo {author} {\bibfnamefont {M.}~\bibnamefont
  {Eisner}},\ }\bibfield  {title} {\enquote {\bibinfo {title} {Radiation from a
  homogeneous isothermal sphere},}\ }\href {\doibase 10.1364/AO.9.002685}
  {\bibfield  {journal} {\bibinfo  {journal} {Appl. Opt.}\ }\textbf {\bibinfo
  {volume} {9}},\ \bibinfo {pages} {2685--2690} (\bibinfo {year}
  {1970})}\BibitemShut {NoStop}%
\bibitem [{\citenamefont {Thompson}\ \emph {et~al.}(2018)\citenamefont
  {Thompson}, \citenamefont {Zhu}, \citenamefont {Mittapally}, \citenamefont
  {Sadat}, \citenamefont {Xing}, \citenamefont {McArdle}, \citenamefont
  {Qazilbash}, \citenamefont {Reddy},\ and\ \citenamefont
  {Meyhofer}}]{thompson2018}%
  \BibitemOpen
  \bibfield  {author} {\bibinfo {author} {\bibfnamefont {Dakotah}\ \bibnamefont
  {Thompson}}, \bibinfo {author} {\bibfnamefont {Linxiao}\ \bibnamefont {Zhu}},
  \bibinfo {author} {\bibfnamefont {Rohith}\ \bibnamefont {Mittapally}},
  \bibinfo {author} {\bibfnamefont {Seid}\ \bibnamefont {Sadat}}, \bibinfo
  {author} {\bibfnamefont {Zhen}\ \bibnamefont {Xing}}, \bibinfo {author}
  {\bibfnamefont {Patrick}\ \bibnamefont {McArdle}}, \bibinfo {author}
  {\bibfnamefont {M~Mumtaz}\ \bibnamefont {Qazilbash}}, \bibinfo {author}
  {\bibfnamefont {Pramod}\ \bibnamefont {Reddy}}, \ and\ \bibinfo {author}
  {\bibfnamefont {Edgar}\ \bibnamefont {Meyhofer}},\ }\bibfield  {title}
  {\enquote {\bibinfo {title} {Hundred-fold enhancement in far-field radiative
  heat transfer over the blackbody limit},}\ }\href {\doibase
  10.1038/s41586-018-0480-9} {\bibfield  {journal} {\bibinfo  {journal}
  {Nature}\ }\textbf {\bibinfo {volume} {561}},\ \bibinfo {pages} {216--221}
  (\bibinfo {year} {2018})}\BibitemShut {NoStop}%
\bibitem [{\citenamefont {Fern\'andez-Hurtado}\ \emph
  {et~al.}(2018)\citenamefont {Fern\'andez-Hurtado}, \citenamefont
  {Fern\'andez-Dom\'{\i}nguez}, \citenamefont {Feist}, \citenamefont
  {Garc\'{\i}a-Vidal},\ and\ \citenamefont {Cuevas}}]{cuevas2018}%
  \BibitemOpen
  \bibfield  {author} {\bibinfo {author} {\bibfnamefont {V.}~\bibnamefont
  {Fern\'andez-Hurtado}}, \bibinfo {author} {\bibfnamefont {A.~I.}\
  \bibnamefont {Fern\'andez-Dom\'{\i}nguez}}, \bibinfo {author} {\bibfnamefont
  {J.}~\bibnamefont {Feist}}, \bibinfo {author} {\bibfnamefont {F.~J.}\
  \bibnamefont {Garc\'{\i}a-Vidal}}, \ and\ \bibinfo {author} {\bibfnamefont
  {J.~C.}\ \bibnamefont {Cuevas}},\ }\bibfield  {title} {\enquote {\bibinfo
  {title} {Super-planckian far-field radiative heat transfer},}\ }\href
  {\doibase 10.1103/PhysRevB.97.045408} {\bibfield  {journal} {\bibinfo
  {journal} {Phys. Rev. B}\ }\textbf {\bibinfo {volume} {97}},\ \bibinfo
  {pages} {045408} (\bibinfo {year} {2018})}\BibitemShut {NoStop}%
\bibitem [{\citenamefont {Patankar}(2018)}]{patankar}%
  \BibitemOpen
  \bibfield  {author} {\bibinfo {author} {\bibfnamefont {Suhas~V}\ \bibnamefont
  {Patankar}},\ }\href {\doibase 10.1201/9781482234213} {\emph {\bibinfo
  {title} {Numerical heat transfer and fluid flow}}}\ (\bibinfo  {publisher}
  {CRC Press},\ \bibinfo {year} {2018})\BibitemShut {NoStop}%
\bibitem [{\citenamefont {Wang}\ and\ \citenamefont {Shi}(2020)}]{WangShi2020}%
  \BibitemOpen
  \bibfield  {author} {\bibinfo {author} {\bibfnamefont {Tian-Shi}\
  \bibnamefont {Wang}}\ and\ \bibinfo {author} {\bibfnamefont {Wan-Yuan}\
  \bibnamefont {Shi}},\ }\bibfield  {title} {\enquote {\bibinfo {title}
  {Transition of marangoni convection instability patterns during evaporation
  of sessile droplet at constant contact line mode},}\ }\href {\doibase
  https://doi.org/10.1016/j.ijheatmasstransfer.2019.119138} {\bibfield
  {journal} {\bibinfo  {journal} {International Journal of Heat and Mass
  Transfer}\ }\textbf {\bibinfo {volume} {148}},\ \bibinfo {pages} {119138}
  (\bibinfo {year} {2020})}\BibitemShut {NoStop}%
\end{thebibliography}%
\end{document}